\documentclass[a4paper,11pt]{article}
\pdfoutput=1 
\usepackage{jcappub} 
\usepackage[T1]{fontenc} 
\usepackage{graphicx}  
\usepackage{todonotes}
\usepackage{changepage}
\usepackage{academicons}
\usepackage{natbib}
\usepackage{multirow}
\usepackage{orcidlink}
\usepackage{subcaption}
\usepackage{academicons}
\usepackage{soul}
\usepackage{xcolor}

\definecolor{orcidlogocol}{HTML}{A6CE39}

\usepackage{aas_macros}

\title{Detecting Axion-like particles using Cosmic Variance Cancellation with CMB and Radio surveys}

\author{
Harsh Mehta$^{a}$\orcidlink{0009-0007-4664-4820},
Anaya Dixit$^{b}$\orcidlink{0009-0007-8123-7796},
Suvodip Mukherjee$^{a}$\orcidlink{0000-0002-3373-5236},
Joseph Silk$^{c,d,e}$\orcidlink{0000-0002-1566-8148}
}

\affiliation{$^{a}$Department of Astronomy and Astrophysics, Tata Institute of Fundamental Research,
Homi Bhabha Road, Mumbai 400005, India}

\affiliation{$^{b}$Indian Institute of Technology, Kharagpur,  Kharagpur 721302, West Bengal, India}

\affiliation{$^{c}$Institut d'Astrophysique de Paris, (UMR7095: CNRS \& UPMC- Sorbonne Universities),
F-75014, Paris, France}

\affiliation{$^{d}$Department of Physics and Astronomy,
The Johns Hopkins University,
Baltimore, MD 21218, USA}
\affiliation{$^{e}$BIPAC, Department of Physics, University of Oxford, Keble Road, Oxford OX1 3RH, UK}

\emailAdd{harsh.mehta@tifr.res.in}
\emailAdd{anaya.dixit@kgpian.iitkgp.ac.in}
\emailAdd{suvodip@tifr.res.in}
\emailAdd{silk@iap.fr}

\begin{document}
\abstract{Axions and axion-like particles (ALPs) arise naturally in many extensions of the Standard Model and are among the well-motivated candidates for dark matter. In the presence of magnetic fields of galaxy clusters, the Cosmic Microwave Background (CMB) photons can convert to ALPs, with the efficiency of the process governed by the cluster electron density and magnetic field profiles, the photon–ALP coupling strength (${g_{a\gamma}}$), as well as the frequency ($\nu$) of the photon at the redshift of the cluster. The CMB blackbody spectrum suggests this resonant conversion takes place at radio wavelengths as well, following the spectral behaviour of the ALP distortion signal. This opens up a new window to search for ALPs using cosmic variance cancellation (CVC), with multi-frequency tracers of the same phenomenon in CMB photon-ALP resonant conversion. The constraints on the ALP signal ratios from different combinations of microwave and radio bands of Simons Observatory (SO) and Square Kilometer Array (SKA), can be significantly improved using CVC as compared to the case of using auto-only spectra from the two experiments. With the large number of galaxy clusters that will be observed by SO and SKA, we will be able to obtain much more information using CVC, especially for the case of low-mass ALPs with stronger signals. Using the auto-only spectra from galaxy clusters up to redshift $z = 1$ for inference of normalized ratio parameter, we obtain a standard deviation of $5.9 \times 10^{-2}$ for ALP mass $m_a = 10^{-14} \, \rm{eV}$, which improves to $1.3 \times 10^{-2}$ using CVC. Not only is this method a universal probe of the ALP distortion signal using its spectral dependence, but will be able to provide a more robust consistency check, helping to identify and mitigate potential spurious signals that might arise in CMB-only analyses, based on its frequency behavior in different bands.

}

\maketitle
\flushbottom

\section{Introduction}
\label{sec:intro}
In the standard $\Lambda$CDM cosmological framework, the energy content of the Universe is partitioned into radiation, matter, and dark energy, each of which governs the cosmic expansion and structure formation during distinct evolutionary epochs \cite{alpher1948evolution,ratra2008beginning,gamow1948evolution}. The matter sector is dominated by dark matter, accounting for approximately $85\%$ of the total matter density, while baryonic matter contributes about $15\%$, with a smaller fraction arising from massive, non-relativistic neutrinos \cite{ade2016planck,2020_planck,aghanim2020planck,Dodelson:2003ft}. 

While the $\Lambda$CDM paradigm has been remarkably successful in describing a wide range of cosmological observations, the fundamental nature of dark matter remains unknown \cite{Dodelson:2003ft,Planck:2015fie,aghanim2020planck,peebles2020large,trimble1987existence,bertone2005particle,bertone2018new,feng2010dark}. This open question has motivated the exploration of extensions to the standard model of particle physics that introduce new, weakly interacting particles beyond the conventional cold dark matter candidates \cite{langacker2017standard,trimble1987existence,bertone2005particle,bertone2018new,feng2010dark,arcadi2018waning,spergel2000observational,dodelson1994sterile,preskill1983cosmology,jungman1996supersymmetric,peccei1977cp}. Among these possibilities, axions and axion-like particles (ALPs) have emerged as particularly well-motivated dark matter candidates, arising naturally in beyond  standard model theories\cite{peccei1977cp,berezhiani1991cosmology,wilczek1978problem}. Originally proposed as a consequence of the Peccei--Quinn mechanism to resolve the strong CP problem in quantum chromodynamics, axions are light, pseudoscalar particles whose properties allow them to be produced abundantly in the early Universe and to behave as cold dark matter on cosmological scales \cite{preskill1983cosmology,abbott1983cosmological,dine1983not}. More generally, ALPs appear in a wide class of scenarios beyond the standard model, and span a broad range of masses and interaction strengths while retaining similar phenomenology \cite{berezhiani1991cosmology,dine1983not,abbott1983cosmological,preskill1983cosmology,Ghosh:2022rta,1992SvJNP..55.1063B,khlopov1999nonlinear,sakharov1994nonhomogeneity,sakharov1996large,chadha2022axion}. Due to their weak couplings to photons and other standard model particles, axions and ALPs are challenging to detect directly, but they can leave subtle imprints on astrophysical and cosmological observables. As a result, the astrophysical systems, along with cosmological probes provide powerful laboratories for probing their existence and constraining their properties.

The radiation component of the Universe is mainly contributed by the Cosmic Microwave Background (CMB), which is the relic radiation from the early Universe. The CMB radiation is observed primarily at microwave wavelengths and is characterized by an almost ideal black-body spectrum \cite{Fixsen_1996,fixsen1996cosmic}. Measurements reveal that the CMB is nearly isotropic across the sky, with temperature fluctuations at the level of only one part in $10^{5}$ relative to the mean temperature \cite{Dodelson:2003ft,Fixsen_1996,fixsen1996cosmic,hanson2009estimators}. This radiation is a fossil remnant of the early Universe, originating from the epoch of last scattering when photons decoupled from the tightly coupled photon--baryon plasma. As these photons propagated through the expanding Universe, their wavelengths were stretched by cosmic expansion, resulting in the present-day black-body temperature of $2.7255\,\mathrm{K}$ \cite{Fixsen_2009}. 

The observed anisotropies in the CMB arise from a combination of physical processes occurring both prior to and after decoupling. Fluctuations generated before decoupling are referred to as primary anisotropies, while those imprinted during the subsequent evolution of the Universe are known as secondary anisotropies \cite{Dodelson:2003ft,hu2002cosmic}. Secondary anisotropies originate from gravitational and electromagnetic interactions between CMB photons and the intervening matter distribution along their paths. In particular, as CMB photons traverse the large-scale structure of the Universe or pass through massive systems such as galaxy clusters, they may experience gravitational lensing due to the cluster potential or interact electromagnetically with baryons. These interactions give rise to a variety of secondary signatures, including effects associated with reionization \cite{adam2016planck}, gravitational lensing \cite{Smith_2007}, and the Sunyaev--Zel'dovich (SZ) effects \cite{1972CoASP...4..173S}. Such effects can be investigated through precise measurements of temperature and polarization fluctuations in the CMB along cluster lines of sight. Moreover, processes such as the thermal-SZ (tSZ) effect modify the black-body spectrum of the CMB itself, producing characteristic spectral distortions that result in deviations in the black-body spectrum \cite{2014PTEP.2014fB107T,erler2018planck}.  

Beyond these well-established mechanisms, additional anisotropic spectral distortions may arise if new light particles beyond the standard model exist. One such possibility involves ALPs, whose production or interaction in astrophysical environments can imprint distinctive signatures on the CMB \cite{dine1983not,abbott1983cosmological,preskill1983cosmology,Ghosh:2022rta,1992SvJNP..55.1063B,khlopov1999nonlinear,sakharov1994nonhomogeneity,sakharov1996large}. In the presence of galaxy clusters, CMB photons propagating through magnetized intracluster media may undergo conversion into ALPs, provided that these particles exist \cite{Mukherjee_2020,Mehta:2024wfo,Mehta:2024pdz}. This process is enabled by a weak axion--photon interaction, parameterized by the coupling constant $g_{a\gamma}$ \cite{Carosi:2013rla,Mukherjee_2020,Ghosh:2023xhs}. When the effective plasma mass of the photon matches the mass of the ALP, resonant conversion can occur, significantly enhancing the probability of photon-ALP mixing and leading to polarized, frequency-dependent distortions in the observed CMB signal along the cluster line of sight. The efficiency of this resonant conversion is governed by several factors, including the strength of the transverse magnetic field, the local electron density at the resonance location, the redshift of the cluster, and the value of the axion-photon coupling $g_{a\gamma}$ \cite{Mukherjee_2018,Mukherjee_2019,Mukherjee_2020,osti_22525054}. With upcoming experiments such as Simons Observatory (SO) \cite{Ade_2019} and LiteBIRD \cite{paoletti2024LiteBIRD,matsumura2014mission,takakura2023wide} , 

we will be able to rule out ALPs of masses $10^{-11} - 10^{-15}$ eV by more than an order of magnitude better than the current bounds from the CERN Axion Solar Telescope (CAST) \cite{2017} using the auto power spectra at microwave wavelengths \cite{Mehta:2024wfo,Mehta:2024pdz,Mehta:2024sye,Mehta:2025slu,mondino2024axioninducedpatchyscreeningcosmic,Goldstein:2024mfp}. The precision of these inferences though, is limited by the cosmic variance of finite sky surveys and observation of a single sky field. {Further information on the resonant conversion phenomenon can be extracted using multiple tracers in the form of multi-band observations in radio, infrared, or X-ray detectors. Not only does such an analysis serve as a universal probe of any new physics but can also check for any {potential contaminants or systematic effects using the spectral behaviour of the ALP signal}. In this work, we explore the idea of probing ALPs using CMB and radio observations.}

 If the CMB photon-ALP resonant conversion occurs, this phenomenon will also show signatures at radio frequencies. In this regime, the frequency dependence of the ALP distortion signal, accompanied by the lowering of the CMB intensity away from its peak wavelength, leads to a weaker ALP distortion signal. These frequencies are already highly impacted by galactic synchrotron emission and other foregrounds, which highly reduce the signal-to-noise ratio in this regime. A way to go about this problem is to suppress the cosmic variance limited observation of the single sky realization, by  using the cross spectra for the ALP signal at both the radio and CMB frequencies, along with the auto power spectra \cite{baleato2023model,sibthorpe2012extragalactic,seljak2009extracting,foreman2019cosmic,schmittfull2018parameter}. Although this would not deliver a significant improvement on the ALP coupling bounds, the correlated signals from the two frequency regimes would allow for a cancellation of the cosmic variance, drastically improving the signal-to-noise ratio on the ratio of the signals at these frequencies. Since the cross spectrum of the foregrounds at these frequencies is weak, and the correlated instrument noise becomes negligible, this method is able to probe the cosmological nature of ALPs using a proxy in terms of these ratios, the inference of which involve minimal systematics. Not only will this technique be able to probe the universal nature of ALPs, {but also provide a consistency check using the spectral nature of the ALP signal.} This technique has been applied in lensing and galaxy clustering analyses, as well as studies on redshift space distortions (RSD) and local non-Gaussianities \cite{baleato2023model,sibthorpe2012extragalactic,seljak2009extracting,foreman2019cosmic,schmittfull2018parameter}. {In this work we apply the idea of cosmic variance cancellation (CVC) to probe the resonant ALP distortion signal with high-resolution CMB and radio surveys.} With the upcoming experiments such as the Square Kilometer Array (SKA) \cite{Carilli_2004,braun2019anticipated} and SO \cite{Ade_2019}, we will be able to resolve around 1 million and 24000 clusters respectively up to a redshift of $z = 3$. The high resolution of SKA of about a few arcseconds \cite{braun2019anticipated} will allow us to probe the ALP signal up to smaller scales or high multipoles. With the high number of spectral bands available with these surveys, we will be able to use information from multiple frequency channels. 

The motivation behind probing the ALP signal across different spectral regimes is discussed in detail in Sec.~\ref{sec:motive}. The ALP distortion signal generated due to the CMB photon-ALP resonant conversion is explained in Sec. \ref{sec:resoconv}. This is followed, in Sec.~\ref{sec:cvc}, by an overview of the cosmic variance cancellation (CVC) technique and the related methodology. The forecasts  obtained from mock observational data using the CVC technique are described   in Sec.~\ref{sec:results}. Finally, Sec.~\ref{sec:conclusion} summarizes the key outcomes of this work and highlights the broader relevance of the CVC technique in the context of upcoming observational facilities. Throughout this study, we adopt natural units ($\hbar = 1$, $c = 1$, $k_B = 1$), unless stated otherwise, and employ the cosmological parameters inferred from the Planck 2018 analysis (TT, TE, EE + lowP + lensing) \cite{aghanim2020planck}.



\section{Motivation}
\label{sec:motive}
A fundamental limitation in extracting cosmological information from observations of the Universe arises from the interplay between instrumental noise and cosmic variance. Instrumental noise originates from the finite sensitivity and resolution of the detectors, introducing statistical uncertainties that can, in principle, be reduced through improved experimental design, longer integration times, or increased sky coverage. In contrast, cosmic variance represents an irreducible source of uncertainty that stems from the fact that we observe only a single realization of the Universe. Even with a perfect, noise-free experiment, fluctuations on large angular scales are fundamentally limited by the finite number of available modes, constraining the precision with which cosmological parameters can be inferred (see Appendix \ref{sec:estimator}) \cite{Dodelson:2003ft,Hu_2002,white1993cosmic,kamionkowski1997getting,colombi2000experimental,szapudi1995cosmic,driver2010quantifying,somerville2004cosmic}. 

As modern cosmological surveys approach the cosmic-variance limit across a wide range of angular scales, further progress increasingly relies on strategies that can mitigate or bypass this fundamental barrier. One such approach is cosmic variance cancellation, which exploits the correlated response of multiple observables to the same underlying large-scale structure. By combining different tracers of the same physical phenomenon or those of the matter distribution, it becomes possible to reduce the impact of cosmic variance and extract additional information beyond what is accessible from any single observable alone. This principle motivates the development of multi-tracer and multi-frequency frameworks that can unlock otherwise inaccessible cosmological information. This concept has been applied in lensing and galaxy clustering analyses, as well as studies concerning RSD and local non-Gaussianities \cite{baleato2023model,sibthorpe2012extragalactic,seljak2009extracting,foreman2019cosmic,schmittfull2018parameter}. 

The precision on the ALP coupling constant using the CMB or other astrophysical probes suffers from the cosmic variance-limited observation \cite{Mehta:2024pdz,Mehta:2024wfo,Mehta:2025slu,mondino2024axioninducedpatchyscreeningcosmic,Goldstein:2024mfp}. Further information on the nature of ALPs can be extracted if this limit is suppressed using multiple tracers of the same resonant phenomenon.  
If the CMB photon-ALP resonant conversion is observable at microwave frequencies, it will also have signatures in the radio regime. Although the signal itself will be very weak at these wavelengths where the foregrounds such as the galactic synchrotron emission \cite{1986rpa..book.....R,syrovatskii1959distribution,haslam1981galactic,davies1996galactic} dominate, combining it with microwave frequencies
will open a new window in being able to probe the ALP signal using CVC. The cross spectra of the two bands sees a weak impact from correlated foregrounds and almost negligible instrument noise. Using this information, one can obtain significant improvements on the constraints on the ratio of the signal in these frequency bands. {Information from other bands, such as X-rays or infrared frequencies, can also be used to obtain better constraints using the CVC technique and can be used to study signals with specific spectral dependencies, but that will be explored in the future.}

\begin{figure}
    \centering    \includegraphics[width=17cm, height=10cm]{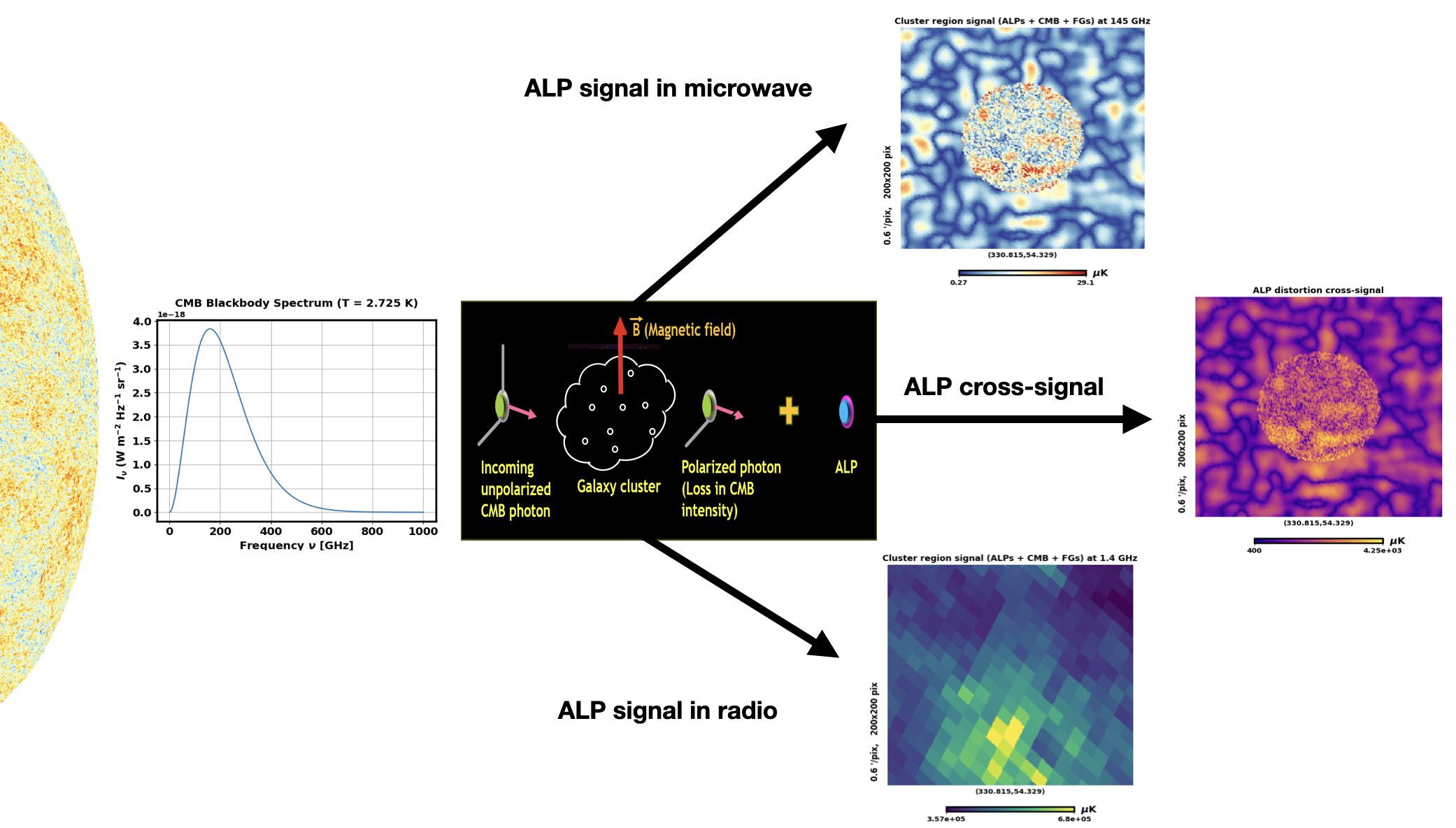}
    \caption{This figure shows how the ALP distortion signal will show up in both the microwave and radio frequencies. 
The CMB photon-ALP resonant conversion will result in a polarized spectral distortion of the CMB blackbody, accompanied with a loss in CMB intensity, and impacting both the radio and microwave frequencies.  
Using the cross CMB-radio observations to probe the ALP signal will provide additional information by drastically improving the signal ratio inferences.}
    \label{fig:cmbradax}
\end{figure}

{In Fig. \ref{fig:cmbradax}, we show how the ALP distortion signal will show up in both the microwave and radio frequencies. 
The CMB photon-ALP resonant conversion will result in a polarized spectral distortion of the CMB blackbody, accompanied with a loss in CMB intensity, and impacting both the radio and microwave frequencies. Using the cross CMB-radio observations to probe the ALP signal will provide additional information by drastically improving the signal ratio inferences. Since the foregrounds and CMB dominate on large scales, it is the small scales around the cluster regions that can be used to probe the ALP signal.}

With the upcoming high-resolution experiments with reduced noise, such as SKA \cite{braun2019anticipated} and SO \cite{Ade_2019}, we can improve drastic improvements in constraints using the CVC technique. Using information from multiple frequency channels from these detectors will not only provide us with better inferences but can also check for {any potential contaminants or systematic effects}, thus probing the universal nature of ALPs by providing smoking-gun signatures in terms of spectral dependence in the two regimes if ALPs exist. This technique can be extended to multi-frequency observations as well using infrared (say, using the Wide-field Infrared Survey Explorer (WISE) \cite{wen2018catalogue}) or X-ray (say, using eROSITA \cite{merloni2012erosita,predehl2021erosita,bulbul2024srgerosita}) surveys. 

\section{CMB Photon-ALP resonant conversion in Galaxy clusters}
\label{sec:resoconv}

\begin{figure}
    \centering    \includegraphics[width=13cm, height=13cm]{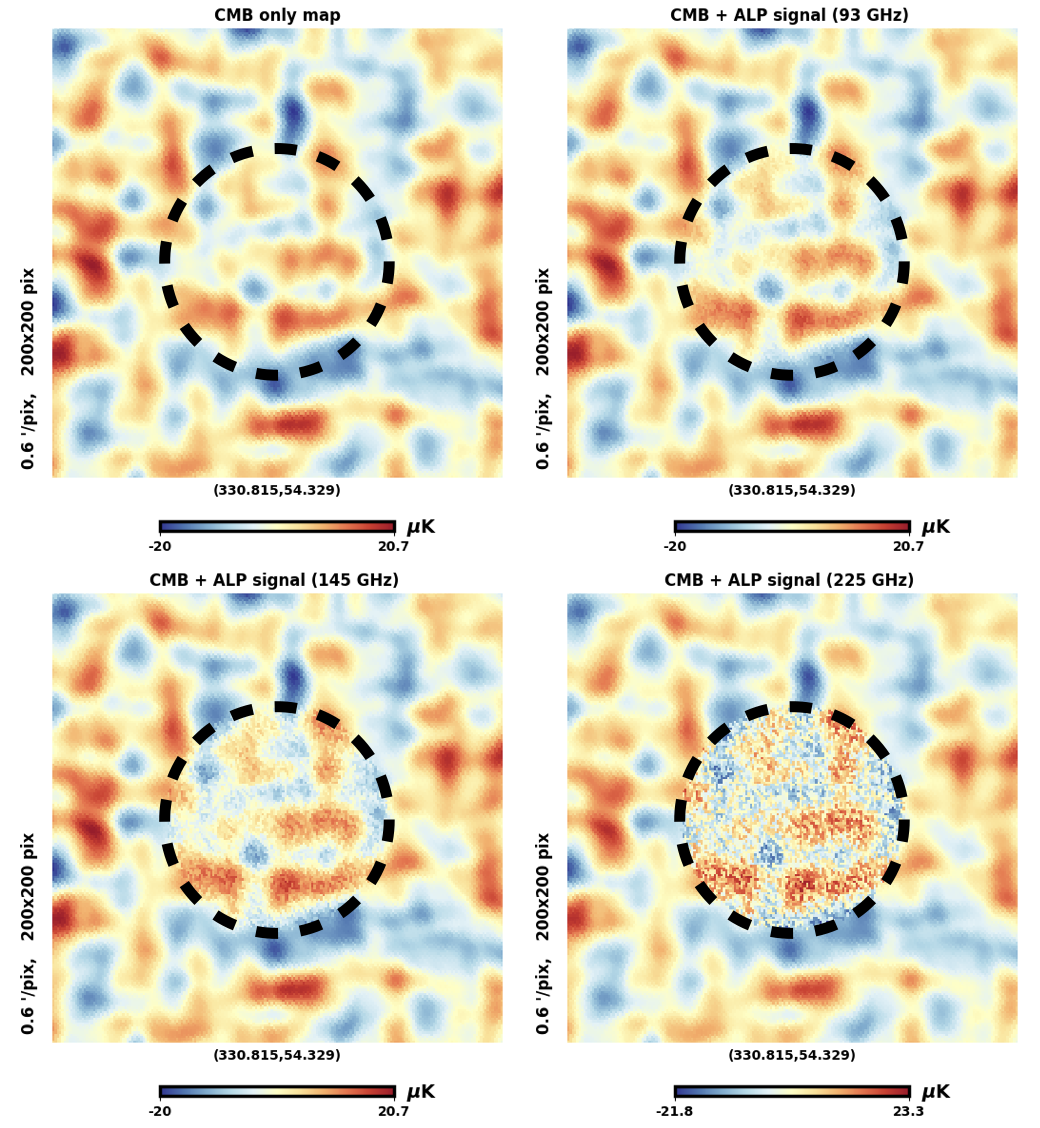}
    \caption{This figure depicts the change that is caused in the CMB polarization map due to the photon-ALP conversion in a cluster (shown as a dashed circle). If there are no ALPs in nature, the fluctuations in the cluster region will be frequency-independent and follow the smooth behaviour of the primordial CMB. But if ALPs exist, there will be additional fluctuations in the cluster region, which will increase with frequency.}
    \label{fig:cmbaxmaps}
\end{figure}
If ALPs exist, photons from the Cosmic Microwave Background propagating through galaxy clusters may undergo conversion into ALPs in the presence of the cluster’s magnetized intracluster medium (ICM). This conversion is mediated by the transverse component of the magnetic field and is governed by the photon–ALP coupling constant $g_{a\gamma}$. Depending on local plasma conditions, the conversion may proceed either resonantly or non-resonantly, with the resonant process typically yielding a significantly enhanced signal. Resonant conversion occurs when the effective photon mass in the plasma is equal to the ALP mass, such that  
\begin{equation}
m_a = m_\gamma = \frac{\hbar \omega_p}{c^2} \simeq \frac{\hbar}{c^2}\sqrt{\frac{n_e e^2}{m_e \epsilon_0}},
\label{eq:resonance mass}
\end{equation}
where $\omega_p$ denotes the plasma frequency and $n_e$ is the local electron number density \cite{Mukherjee_2020}. For typical electron densities encountered in galaxy clusters, this condition allows sensitivity to ALPs in the mass range $m_a \sim 10^{-15}$--$10^{-11}$~eV.

The interaction between photons and ALPs is described by the effective Lagrangian \cite{Raffelt:1996wa}
\begin{equation}
\mathcal{L}_{\mathrm{int}} = -\frac{g_{a\gamma}}{4} F_{\mu\nu} \tilde{F}^{\mu\nu} a 
= g_{a\gamma} \, \mathbf{E}\cdot\mathbf{B}_{\mathrm{ext}}\, a ,
\label{eq:lagr}
\end{equation}
where $F_{\mu\nu}$ and $\tilde{F}^{\mu\nu}$ denote the electromagnetic field tensor and its dual, respectively, and $a$ represents the ALP field. This interaction explicitly requires a magnetic field component transverse to the photon propagation direction, implying that photon–ALP conversion is suppressed in regions where only longitudinal magnetic fields are present.

To quantify the conversion process, it is convenient to define the relevant mixing parameters that govern photon–ALP oscillations. These include the effective mass terms and the mixing amplitude,
\begin{equation}
\label{eq:lenparams}
\begin{split}
\Delta_a &= -\frac{m_a^2}{2\omega}, \qquad
\Delta_e \simeq -\frac{\omega_p^2}{2\omega}, \\
\Delta_{a\gamma} &= \frac{g_{a\gamma} B_t}{2}, \qquad
\Delta_{\mathrm{osc}}^2 = (\Delta_a - \Delta_e)^2 + 4\Delta_{a\gamma}^2 ,
\end{split}
\end{equation}
where $B_t$ is the transverse magnetic field strength and $\omega$ is the photon angular frequency. 

The conversion efficiency is governed by the adiabaticity parameter $\gamma_{\mathrm{ad}}$, which compares the oscillation length of the photon–ALP system to the spatial scale on which the plasma properties vary. In the adiabatic regime ($\gamma_{\mathrm{ad}} \gg 1$), conversion is highly efficient, while in the opposite non-adiabatic limit ($\gamma_{\mathrm{ad}} \ll 1$), the transition probability is suppressed. In the latter case, the conversion probability is well approximated by the Landau–Zener formula \cite{Mukherjee_2020,Mehta:2024wfo,osti_22525054}
\begin{equation}
P(\gamma \rightarrow a) = 1 - e^{-\pi \gamma_{\mathrm{ad}}/2} \simeq \frac{\pi}{2}\gamma_{\mathrm{ad}} ,
\label{eq:prob}
\end{equation}
where the adiabaticity parameter is given by
\begin{equation}
\gamma_{\mathrm{ad}} = \frac{\Delta_{\mathrm{osc}}^2}{|\nabla \Delta_e|}
= \left| \frac{2 g_{a\gamma}^2 B_t^2 \nu (1+z)}{\nabla \omega_p^2} \right|.
\label{eq:gamma_ad}
\end{equation}
Here, $\nu(1+z)$ denotes the photon frequency at the cluster redshift.
The CMB intensity changes along the cluster line of sight and is given as:
\begin{equation}
    {
    \Delta I^{\rm{ax}} \approx \pi \gamma_{ad} \left( \frac{2h \nu^3}{c^2}  \right) \frac{1}{e^{h\nu /k_B T_{\rm{cmb}}} - 1} .}
    \label{eq:Distort}
\end{equation}
In the low frequency limit applicable for microwave and radio frequencies, we can use the Rayleigh-Jeans (RJ) approximation and convert intensity units to RJ temperature units. This conversion goes as:
\begin{equation}
    \label{eq:rj}
    \Delta T_{RJ} = \frac{c^2}{2k_{B} \nu^2} \Delta I.
\end{equation}
The ALP distortion signal in the RJ temperature units thus varies as, 
\begin{equation}
    \label{eq:rj}
    \Delta T_{RJ} \propto \nu \, g_{a\gamma}^2.
\end{equation}
We will use these units in our analysis. The dependence on $\nu$ will allow us to use the CVC technique to obtain improved constraints on the ratios of the signals at different frequencies.

{For propagation paths through the intracluster medium, the polarization plane of CMB photons is expected to experience Faraday rotation due to the presence of a magnetic field within an ionized medium, which influences the motion of charged particles, causing electrons to spiral preferentially in a direction determined by the magnetic field orientation. A linearly polarized electromagnetic wave can be expressed as a superposition of left- and right-handed circularly polarized modes. When such radiation propagates through a magnetized plasma, these two circular polarization components experience different refractive indices. This difference introduces a relative phase shift between them, which results in a rotation of the net linear polarization angle \cite{ghatak2009optics,faraday1839experimental,gardner1966polarization,burn1966depolarization}. The change in  polarization angle is proportional to the square of the photon wavelength and can be written as $\Delta \chi = RM \lambda^2$, where $\lambda$ denotes the wavelength of the radiation. The Rotation Measure ($RM$) quantifies the cumulative effect of the magnetized plasma along the propagation path and is defined as \cite{burn1966depolarization,GOVONI_2004,murgia2004magnetic,bohringer2016cosmic,Ferrari_2008,clarke2001new,eilek2002magnetic,bonafede2010galaxy,gardner1966polarization}:

\begin{equation}
RM = 811.9 \int_{0}^{L} n_e B_{||} \, dl \; \mathrm{rad \, m^{-2}} .
\label{eq:FR}
\end{equation}
Here $n_e$ is the electron density and $B_{||}$ is the longitudinal component of magnetic field, the product of which is integrated along the line of sight.
For typical propagation paths through a galaxy cluster, Faraday rotations of approximately an arcminute at frequencies $\nu \geq 90$ GHz are expected. Since this angular rotation is below the characteristic resolution of the instrument beam of experiments such as SO, its observational impact is minimal. For radio frequencies, this will be a major contributor to the auto-power spectra (which are already highly contaminated by foregrounds), and its contribution can be inferred using radio polarization observations of galaxy clusters \cite{murgia2004magnetic,carilli2002cluster,GOVONI_2004,prestage1988cluster}. 
In our analysis, we deal with the polarization intensity, which is given as: 
\begin{equation}
    \Delta T = \sqrt{\Delta T_{Q}^2 + \Delta T_{U}^2},
    \label{eq:polint}
\end{equation}
where $T_{Q}$ and $T_{U}$ are the polarization components in  the $Q$ and $U$ maps, respectively \cite{ghatak2009optics,Austermann_2012,2020_planck}.
Using the intensity values, the individual polarization components are not  needed to model the ALP signal. Accordingly, Faraday rotation is neglected in the present analysis.}

In an idealized scenario involving a spherically symmetric galaxy cluster with smooth and homogeneous electron density and magnetic field profiles, ALPs of a given mass are produced within a thin spherical shell where the resonance condition is satisfied. When projected onto the plane of the sky, this shell manifests itself as a circular feature, with the conversion probability peaking near the shell boundary and vanishing outside it. Along any given line of sight, such a configuration admits at most two resonance points, yielding a single effective conversion event \cite{Mukherjee_2020}. {The presence of turbulence in galaxy clusters \cite{xu2009turbulence,dolag2005turbulent,schekochihin2006turbulence,vazza2011massive,schuecker2004probing,sunyaev2003detectability,subramanian2006evolving,brunetti2007compressible} will impact the electron density and magnetic field profiles, hence affecting the ALP signal and will generate non-Gaussian signal \cite{Mehta:2025qfr}. Turbulence in magnetic field may lead to depolarization of the ALP signal, as well as additional fluctuations in the ALP signal, while turbulence in electron density will change the signal  disk size as outer regions of the cluster contribute to the ALP distortion  signal. We consider homogeneous and smooth profiles in this work as turbulence in clusters will not affect the spectral dependence of the ALP signal, hence the CVC technique using multiple frequency bands will still be applicable}. 

Polarization observations of the cosmic microwave background (CMB) provide a powerful avenue to search for axion-like particle (ALP) signatures along the line of sight to galaxy clusters. This is illustrated in Fig. \ref{fig:cmbaxmaps}. For clusters that are spatially resolved, the imprint induced by ALP conversions is expected to appear as a disk-like feature, whose angular extent is set by the mass of the produced ALPs. The resulting distortions manifest as enhanced small-scale fluctuations within the cluster region that grow in amplitude with increasing microwave frequency. In contrast, the CMB signal around the cluster retains its primordial smoothness, exhibiting fluctuations that are largely frequency-independent \cite{Mehta:2024wfo,Mehta:2024sye}.

The spectral dependence ($\Delta T^{\rm{ax}} \propto \nu$) of the ALP distortion signal will enable us to apply the CVC technique using CMB and radio surveys, as the same resonant phenomenon is traced by the distortion signal at microwave as well as radio frequencies. This spectral signature not only allows us to clean the ALP distortion signal, but also provides a universal probe for the detection of ALPs. In the next section, we look at this concept in detail, and apply it for the configurations of upcoming detectors.

\section{Cosmic Variance Cancellation on the ALP signal in the joint CMB and Radio Analysis}

\label{sec:cvc}

The ability to overcome the limitations imposed by cosmic variance is central to extracting maximal information from cosmological observations. While instrumental noise can, in principle, be mitigated through improved detector sensitivity, longer integration times, or increased sky coverage, cosmic variance represents a more fundamental constraint arising from the finite number of observable modes in our Universe \cite{Dodelson:2003ft,white1993cosmic,kamionkowski1997getting,colombi2000experimental,szapudi1995cosmic,driver2010quantifying,somerville2004cosmic}. As a result, even ideal experiments become sample-variance limited, preventing further reduction in statistical uncertainties through conventional observational strategies alone.

The concept of cosmic variance cancellation (CVC) provides a powerful framework for addressing this challenge. Rather than attempting to suppress variance through improved measurements of a single observable, CVC exploits the fact that different cosmological observables often trace the same underlying matter fluctuations. When two or more fields respond coherently to the same large-scale structure, their correlated fluctuations can be combined in such a way that the shared cosmic variance cancels out, leaving behind a cleaner measurement of the underlying physical parameters \cite{baleato2023model,sibthorpe2012extragalactic,seljak2009extracting,foreman2019cosmic,schmittfull2018parameter}. This approach effectively transforms the cosmic variance from a fundamental limitation into a resource that can be leveraged to extract additional information.
\begin{figure}
    \centering    \includegraphics[width=17cm, height=6.5cm]{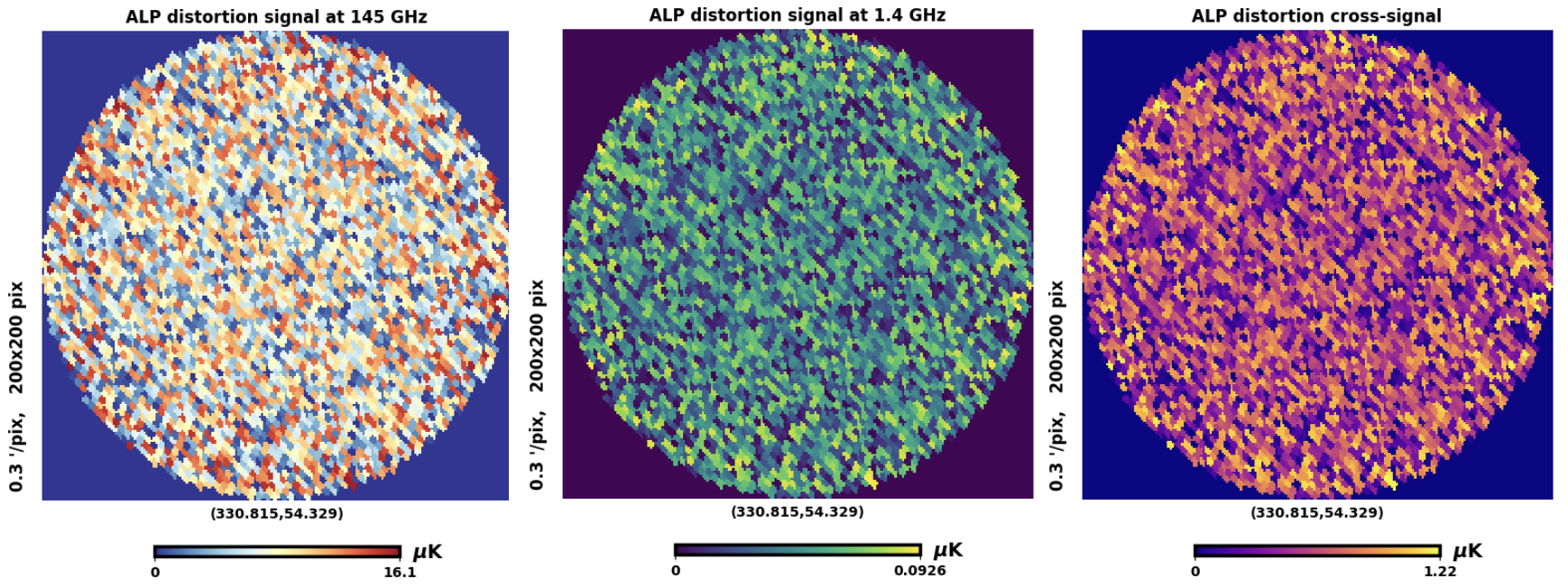}
    \caption{This figure shows  how the variation of ALP distortion intensity fluctuations in a cluster region with frequency changes in magnitude, but the pattern of the ALP distortion fluctuations is similar in the cases of microwave (145 GHz) and radio (1.4 GHz) signals. The hot (cold) regions in the microwave band appear as hot (cold) regions in the radio band as well.  When the cross-signal of the two bands is taken, i.e, $\Delta T_{\nu_1,\nu_2} = \sqrt{\Delta T_{\nu_1} \Delta T_{\nu_2}}$, we see that the signal fluctuations still show a similar pattern as in the microwave and radio maps,   hot (cold) fluctuations show up as hot (cold) fluctuations in the cross signal.} 
    \label{fig:cvcfluc}
\end{figure}

In Fig. \ref{fig:cvcfluc}, we show how the variation of ALP distortion intensity fluctuations with frequency changes in magnitude, but the pattern of the ALP distortion fluctuations is similar in the cases of microwave (145 GHz) and radio (1.4 GHz) signals. {The hot (cold) regions in the microwave band appear as hot (cold) regions in the radio band as well. When the cross-signal of the two bands is taken, i.e, $\Delta T_{\nu_1,\nu_2} = \sqrt{\Delta T_{\nu_1} \Delta T_{\nu_2}}$, we see that the signal fluctuations still show a similar pattern as in the microwave and radio maps,   hot (cold) fluctuations show up as hot (cold) fluctuations in the cross signal.} This pattern of fluctuations is what CVC uses to improve inferences on the signal ratios.

\begin{figure}
    \centering    \includegraphics[width=13cm, height=8cm]{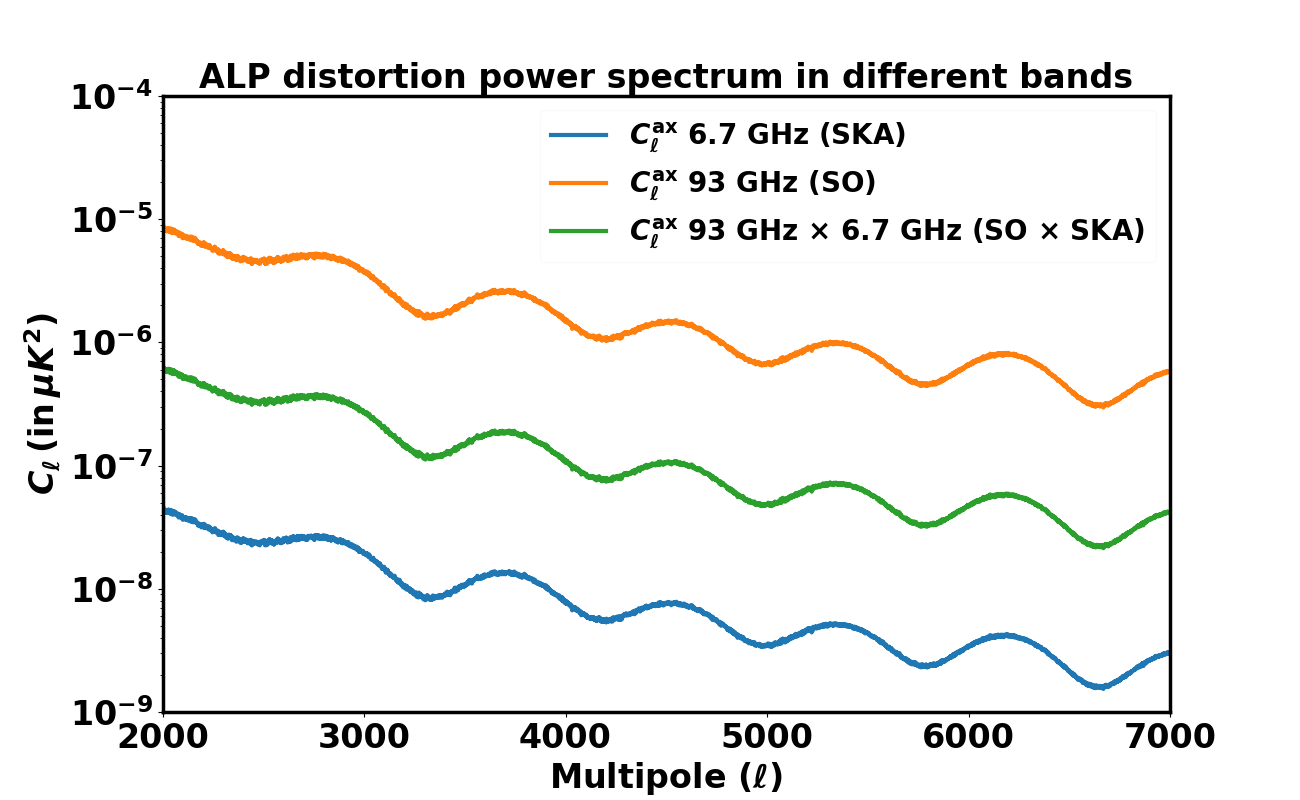}
    \caption{This figure depicts the power spectrum of the auto and cross spectra for the frequency bands 93 GHz from SO and 6.7 GHz from SKA. The modeled ALP power spectra are shown for $g_{a\gamma} = 3 \times 10^{-12} \, \rm{GeV}^{-1}$.  The CVC uses the information that the information that the tracers at different frequencies show similar fluctuations, improving the cross-signal inferences. }
    \label{fig:cvcfig}
\end{figure}

 To explain CVC, it is also necessary to look at the auto and cross-power spectra for the microwave and radio bands.
In Fig. \ref{fig:cvcfig}, we show the variation of the auto and cross power spectra for the frequency bands 93 GHz from SO and 6.7 GHz from SKA for SO observable clusters in the redshift bin $0.3 \leq z < 0.4$. The modeled ALP power spectra are shown for $g_{a\gamma} = 3 \times 10^{-12} \, \rm{GeV}^{-1}$. 
Observations in the microwave regime will place constraints on the ALP coupling constant and the signal ratios, the precision of  which will be limited by the cosmic variance. The CVC technique drastically improves the constraints on signal ratio using cross-spectra.
The ALP distortion power spectra vary with frequency but show similar oscillatory pattern at these wavelengths. CVC uses the information that the tracers at different frequencies show similar spatial fluctuations, improving the cross-signal inferences, as the cross signals which are devoid of substantial noise and foreground contamination improve the signal to noise ratio significantly for the signal ratio parameters for the two frequencies. 

{In this analysis, we approximate both the SO and SKA datasets as beam-smoothed maps, which serves as a convenient first-order framework for estimating the potential of cosmic variance cancellation. This simplification does not fully capture the observational realities of SKA though. As an interferometric array, SKA measures visibilities rather than direct sky maps, and its inherently incomplete and anisotropic uv-coverage introduces a transfer function that varies with both angular scale and direction \cite{Carilli_2004, braun2019anticipated, subrahmanyan2004radio}. This response can attenuate or entirely remove specific spatial modes, with large-scale modes being particularly affected.
By contrast, SO reconstructs maps through a scanning strategy that imposes a distinct form of filtering and mode selection \cite{Ade_2019, hanany2012cmb}. Because cosmic variance cancellation depends critically on the degree of overlap in the recovered modes between different datasets, discrepancies in their respective transfer functions and mode sensitivities can limit the effectiveness of this technique relative to the idealized scenario assumed here.
A more realistic implementation would therefore require explicitly accounting for the interferometric measurement process and imaging pipeline of SKA, including its uv-sampling and map-making procedures. Incorporating these effects is essential for applying cosmic variance cancellation to actual observations, and we defer such a treatment to future work.}

\subsection{Basic formalism of the Cosmic Variance Cancellation}

We consider two statistically isotropic Gaussian fields in the sky,
denoted by $a$ and $b$, with spherical-harmonic coefficients
$a_{\ell m}$ and $b_{\ell m}$.
Their auto and cross-power spectra are defined as
\begin{align}
\langle a_{\ell m}^* a_{\ell m} \rangle &= C_\ell^{aa}; \\
\langle b_{\ell m}^* b_{\ell m} \rangle &= C_\ell^{bb}; \\
\langle a_{\ell m}^* b_{\ell m} \rangle &= C_\ell^{ab}.
\end{align}

\subsection{Inference on the ratio parameter}

We define the ratio parameter
\begin{equation}
\xi = \sqrt{\frac{C_\ell^{bb}}{C_\ell^{aa}}},
\end{equation}
which measures the relative response of field $b$ to the same underlying
phenomenon traced by $a$. We hereby infer the variance on $\xi$ for the cases of auto-spectra only and CVC analyses. { The parameter $\xi$ not only compares the relative powers in various tracers, but can also be used as smoking-gun signature of new physics based on its spectral dependence or effect in different tracers. } 

\subsubsection*{Auto-spectra only:}
We estimate $\xi$ using only the auto-spectra
$C_\ell^{aa}$ and $C_\ell^{bb}$.
Assuming Gaussian statistics, their variances are \cite{Dodelson:2003ft,Hu_2002}:
\begin{align}
\mathrm{Var}(\hat C_\ell^{aa}) &= \frac{2}{2\ell+1} (C_\ell^{aa} )^2, \\
\mathrm{Var}(\hat C_\ell^{bb}) &= \frac{2}{2\ell+1} (C_\ell^{bb}  )^2 .
\end{align}
Using error propagation for
\[
\xi = \sqrt{\frac{C_\ell^{bb}}{C_\ell^{aa}}},
\]
we obtain
\begin{align}
\mathrm{Var}(\xi)
&= \left(\frac{\partial \xi}{\partial C_\ell^{aa}}\right)^2
\mathrm{Var}(C_\ell^{aa})
+
\left(\frac{\partial \xi}{\partial C_\ell^{bb}}\right)^2
\mathrm{Var}(C_\ell^{bb}) \\
&= \frac{1}{4\xi^2}
\left[
\frac{2(C_\ell^{bb})^2}{(2\ell+1)(C_\ell^{aa})^2}
+
\frac{2(C_\ell^{bb})^2}{(2\ell+1)(C_\ell^{aa})^2}
\right].
\end{align}
Using $C_\ell^{bb}=\xi^2 C_\ell^{aa}$, this simplifies to
\begin{equation}
(\sigma_\xi^2)_{\rm{auto}}
=
\frac{\xi^2}{2\ell+1}.
\end{equation}
This uncertainty is entirely dominated by cosmic variance.

\subsubsection*{}

Now we consider the full data vector with auto and cross spectra:
\[
\mathbf{x}_{\ell m} =
\begin{pmatrix}
a_{\ell m} \\
b_{\ell m}
\end{pmatrix},
\qquad
\mathbf{C}_\ell =
\begin{pmatrix}
C_\ell^{aa} + N_\ell^{aa} & C_\ell^{ab} \\
C_\ell^{ab} & C_\ell^{bb} + N_\ell^{bb}
\end{pmatrix}.
\]
The likelihood for $\mathbf{C}_\ell$ is Wishart, and the Fisher information
for a parameter $\theta$ is
\begin{equation}
F_{\theta\theta}
=
\frac{1}{2}(2\ell+1)
\mathrm{Tr}
\left[
\mathbf C_\ell^{-1}
\frac{\partial \mathbf C_\ell}{\partial \theta}
\mathbf C_\ell^{-1}
\frac{\partial \mathbf C_\ell}{\partial \theta}
\right].
\end{equation}
For the parameter $\xi$, we have
\[
\frac{\partial \mathbf C_\ell}{\partial \xi}
=
\begin{pmatrix}
0 & C_\ell^{aa} \\
C_\ell^{aa} & 2\xi C_\ell^{aa}
\end{pmatrix}.
\]
Evaluating the Fisher matrix and taking the inverse of it yields:
\begin{equation}
(\sigma_\xi^2)_{\rm{CVC}}
=
\frac{1}{2\ell+1}
\left(
\frac{N_\ell^{bb}}{C_\ell^{aa}}
+
\xi^2 \frac{N_\ell^{aa}}{C_\ell^{aa}}
\right),
\end{equation}
where $N_\ell^{aa}$ and $N_\ell^{bb}$ are the instrumental noise spectra.
Crucially, in the signal-dominated limit
$
C_\ell^{aa} \gg N_\ell^{aa}, N_\ell^{bb}
$,
the variance tends to zero:
\[
(\sigma_\xi^2)_{\rm{CVC}} \rightarrow 0 \, .
\]

This is the essence of cosmic variance cancellation. If two or more different tracers probe the same underlying density field, their ratio can be measured much more precisely than either alone, because the cosmic variance cancels out. Instrument noise remains, but the sample variance drops out. {If the fields follow a certain spectral dependence, information in multiple frequency bands can be used to obtain better constraints. Thus, this technique can be used as a universal probe for new signals.

In the limit that one of the tracers is signal-dominated, while the other is noise-dominated, i.e., when $
C_\ell^{aa} \gg N_\ell^{aa}$, but $ C_\ell^{bb} \ll  N_\ell^{bb}
$,
 the covariance depends on the relative ratios of $N_{\ell}^{bb}$ and $C_{\ell}^{aa}$:
 \[
(\sigma_{\xi}^2)_{\rm{CVC}} \rightarrow \frac{1}{2\ell + 1}\frac{N_{\ell}^{bb}}{C_{\ell}^{aa}} \, .
\]
 For low SKA noise levels and low mass ALPs, we expect the ALP signal to be strong and detectable in SO, leading to a strong $C_{\ell}^{aa}$ component and hence the covariance will weaken. }

\subsection{Inference on global parameters}

Now consider a different parameterization:
\begin{equation}
C_\ell^{aa} = g S_\ell^{aa}, \quad
C_\ell^{bb} = g S_\ell^{bb}, \quad
C_\ell^{ab} = g S_\ell^{ab},
\end{equation}
where $g$ is a single overall amplitude.

In this case,
\[
\frac{\partial \mathbf C_\ell}{\partial g} = \mathbf S_\ell,
\]
and the Fisher information becomes
\begin{equation}
F_{gg}
=
\frac{1}{2}(2\ell+1)
\mathrm{Tr}
\left[
(\mathbf C_\ell^{-1}\mathbf S_\ell)^2
\right].
\end{equation}
Since all spectra scale identically with $g$, the covariance matrix has
only one effective mode sensitive to $g$.
Consequently,
\[
{\sigma_{g}^2}
=
\frac{2 g^2}{2\ell+1}
\]
independent of whether cross-spectra are included.

From a statistical perspective, cosmic variance cancellation relies on constructing joint estimators that incorporate multiple tracers of the same underlying density field. These tracers may differ in their bias, frequency dependence, or physical origin, but remain correlated through their common dependence on the same phenomenon or the same matter distribution. By appropriately combining their respective power spectra and cross-correlations, it becomes possible to isolate parameters of interest with uncertainties that scale more favorably than those obtained from any single tracer alone. In this sense, the multi-tracer approach generalizes traditional power spectrum analyses by explicitly exploiting cross-covariances between observables rather than treating them as independent sources of noise \cite{baleato2023model,sibthorpe2012extragalactic,seljak2009extracting,foreman2019cosmic,schmittfull2018parameter}.

Within this framework, the formalism naturally extends to the use of multi-frequency and multi-component data sets, where each observable carries distinct instrumental and astrophysical systematics. The key requirement is that these observables probe the same underlying modes, enabling partial or complete cancellation of cosmic variance at the level of the likelihood. This principle has been successfully applied in studies of galaxy clustering, RSD, primordial non-Gaussianity, and CMB \cite{baleato2023model,sibthorpe2012extragalactic,seljak2009extracting,foreman2019cosmic,schmittfull2018parameter}.

\subsection{Methodology}
\begin{table}[htbp]
  \small
  \centering
  \subfloat[\textbf{Simons observatory}]{%
    \hspace{0.5cm}%
    \begin{tabular}{|c|c|c|}
        \hline
        $\nu$ (in GHz) & Noise (in $\mu$K-arcmin) & Beam (arcmin) \\
        \hline
        \hline
        93 & 5.8 & 2.2 \\
        \hline
        145 & 6.3 & 1.4 \\
        \hline
        225 & 15 & 1.0 \\
        \hline
    \end{tabular}%
    \hspace{.5cm}%
  }\hspace{0.01cm}
  \subfloat[\textbf{Square Kilometer Array}]{%
    \hspace{0.5cm}%
    \begin{tabular}{|c|c|c|}
        \hline
        $\nu$ (in GHz) & Noise (in $\mu$Jy/beam/hr) & Beam (in arcsec) \\
        \hline
        \hline
        0.3 & 14 & 40 \\
        \hline
        0.77 & 4.4 & 30 \\
        \hline
        1.4 & 2.2 & 10 \\
        \hline
        6.7 & 1.3 & 8 \\
        \hline
        12.5 & 1.2 & 7 \\
        \hline
    \end{tabular}%
    \hspace{.5cm}%
  }
  \caption{\textbf{Instrument noise and beams for the upcoming experiments SO (CMB survey) \cite{Ade_2019} and SKA (radio survey) \cite{braun2019anticipated}} }
  \label{tab:noise}
\end{table}

We apply the CVC technique to probe the ALP distortion signal using a combination of CMB and radio surveys. 
We use multiple frequency channels from the two surveys: three from SO (93, 145 and 225 GHz channels) and five from SKA (0.3, 0.77, 1.4, 6.7 and 12.5 GHz channels) with their corresponding noise and beam resolution as mentioned in Table \ref{tab:noise}. The SKA noises are integrated over 10 days of observation time. Since the CMB polarization power spectrum can be used to probe the ALP distortion signal, we use the SO noises corresponding to the polarization $Q$ and $U$ maps. We perform our analysis using polarized intensity (see Eq. \eqref{eq:polint}) values, so we do not need the individual components to model the ALP signal.

As shown in Fig. \ref{fig:cmbaxmaps}, galaxy clusters subtend relatively small angular scales, corresponding to the high–multipole regime of the angular power spectrum. It is in this small-scale regime that the signatures of ALP-induced distortions are expected to be most prominent. Consequently, extracting meaningful constraints on the ALP signal requires an accurate characterization of the power spectrum within these localized cluster regions.
To perform this analysis, we employ the \texttt{HEALPy} package \cite{Zonca2019,2005ApJ...622..759G}, which provides an efficient estimation of the power spectrum in masked regions.
Also, before estimating the power spectra, we convert all of the maps to Rayleigh-Jeans (RJ) temperature units in Kelvin.  This assigns a specific frequency dependence to the ALP distortion signal $\Delta T^{\rm{ax}} \propto \nu \, g_{a\gamma}^2$ (see Sec. \ref{sec:resoconv}). Hence, all ratios of the signals at different frequencies will depend on this proportionality.

\subsubsection{Modeling Galaxy clusters at different redshifts}\label{sec:makemap}

\begin{figure}
    \centering    \includegraphics[width=10cm, height=6.5cm]{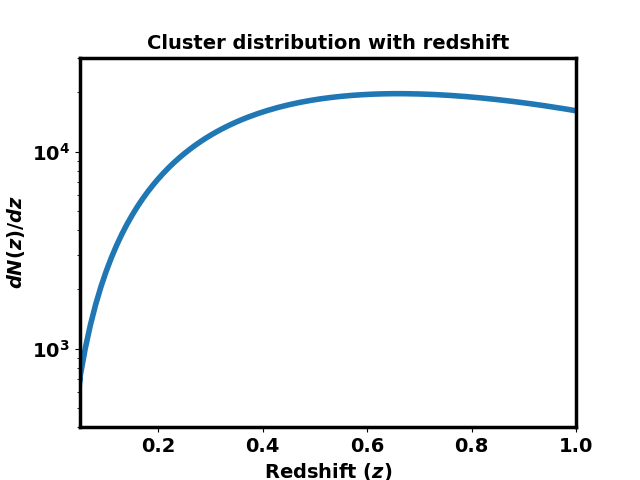}
    \caption{This figure shows  the distribution of galaxy clusters across redshifts that will be observed by SO.} 
    \label{fig:dndz}
\end{figure}
SKA is expected to observe around 1 million clusters up to a redshift of $z=3$, while SO will be able to observe around 24000 clusters up to the same redshift \cite{Carilli_2004,braun2019anticipated,Ade_2019}. We assume that low-redshift clusters up to redshift $z = 1$ detectable by SO, will be resolvable by SKA as well. Thus, we calculate the number of clusters at various redshifts  We generate galaxy clusters in bins of $\Delta z = 0.1$ up to redshift $z = 1$. 

We adopt a sky coverage of 50\% by masking the Galactic plane in order to mitigate contamination from diffuse Galactic foregrounds. Within this observed region, we generate mock realizations of galaxy clusters distributed across redshift. For each cluster, the relevant astrophysical properties—namely the electron density profile and the magnetic field configuration—are randomly drawn from physically motivated ranges. These profiles are described in Appendix \ref{sec:profiles}. 
This approach allows us to capture the intrinsic diversity of cluster environments while remaining consistent with current observational constraints.

The total number of clusters contributing to the signal is determined by integrating the redshift-dependent cluster abundance. We model this distribution using a phenomenological form motivated by observational studies, characterized by a scale parameter $z_0 = 0.33$, corresponding to a mean redshift of $\langle z \rangle \simeq 1$. The differential number of clusters as a function of redshift is given by \cite{Ade_2019}
\begin{equation}
    {\frac{dN_g}{dz} = \frac{z^2}{2 z_0^3} \exp\left(-\frac{z}{z_0}\right)} \, .
\end{equation}
{The distribution is shown in Fig. \ref{fig:dndz}, and shows a steep increase followed by a decay in the distribution at higher redshifts due to observational biases, as well as the limited resolution of the instrument.} 
Using this prescription, we compute the expected number of clusters within each redshift bin and generate the corresponding cluster population. The resulting redshift distribution adopted in our analysis is summarized in Table~\ref{tab:dndz}.

\begin{table}[htbp]
  \small
  \centering
  \caption{\textbf{Number of clusters in various redshift bins} }
  \label{tab:dndz}
  \subfloat{
    \begin{tabular}{|c|c|}
        \hline
         Redshift bins & No. of clusters \\
        \hline
        \hline
        0.05 $\leq z <$ 0.1 & 79  \\
        \hline
         0.1 $\leq z <$ 0.2 & 487\\
        \hline
         0.2 $\leq z <$ 0.3 & 969 \\
        \hline
         0.3 $\leq z <$ 0.4 & 1400\\
        \hline
         0.4 $\leq z <$ 0.5 & 1712 \\
        \hline
         0.5 $\leq z <$ 0.6 & 1892 \\
        \hline
         0.6 $\leq z <$ 0.7 & 1956 \\
        \hline
         0.7 $\leq z <$ 0.8 & 1926 \\
        \hline
         0.8 $\leq z <$ 0.9 & 1830 \\
        \hline
         0.9 $\leq z <$ 1 & 1690 \\
        \hline
    \end{tabular}
    }

\end{table}

To isolate the cluster regions, we apply a spatial mask defined as a step-function window: pixels corresponding to cluster locations are assigned a value of unity, while all other sky pixels are set to zero. The resulting masked map is then used to compute the angular power spectrum in the following sections.

\subsubsection{Modeling of the ALP power   spectrum from Galaxy clusters}\label{sec:makemap}

We construct the ALP-induced signal in galaxy clusters using the formalism described in Sec.~\ref{sec:resoconv}, evaluating the expected spectral distortion at radio and microwave frequencies. For the purpose of modeling, we adopt a standard ALP-photon coupling strength of 
\[
g_{a\gamma} = g_{0} = 10^{-12}\,\mathrm{GeV^{-1}},
\]
which serves as a reference value for generating the theoretical ALP-induced signal.  
We consider ALPs in the mass range $m_a \sim 10^{-15} - 10^{-11}\,\mathrm{eV}$, motivated by the range of electron densities typically encountered in galaxy clusters at low redshifts. For each assumed ALP mass, the corresponding resonant conversion condition determines the spatial region within the cluster where the signal is generated. Using the cluster electron density and magnetic field profiles, the ALP-induced distortion is computed independently for each mass value.
The resulting ALP distortion maps are then used to construct angular power spectra in different redshift bins. 

This power spectrum estimator for two fields $a$ and $b$ is calculated as \cite{Dodelson:2003ft,Hu_2002}:
\begin{equation}
(C_{\ell}^{ab}) = \frac{1}{2\ell + 1}\sum_{m=-\ell}^{\ell} a_{\ell m}^*b_{\ell m} \, .
\label{eq:clab}
\end{equation}
This follows from the isotropy of two fields $a$ and $b$ and the power spectrum estimator formalism, which is explained in Appendix \ref{sec:estimator}, with the spherical harmonic coefficients for the fields being $a_{\ell m}$ and $b_{\ell m}$ respectively. 
 There will be mode-coupling induced due to masking and partial sky observations, which leads to power mixing among different multipoles, hence affecting the power spectrum. This effect is explained in Appendix \ref{sec:partial_sky} and is incorporated by computing the power spectrum of the masked map using the \texttt{anafast} routine in \texttt{HEALPy}.
 
We evaluate the auto-power spectra of the ALP distortion maps for all the 8 frequency channels (3 from SO and 5 from SKA), while the cross-power spectra are evaluated for all possible combinations including one frequency channel from SO and one frequency channel from SKA. This gives us 15 cross-spectra, which will then be used to obtain constraints on the ratio of the signals at different frequencies.  
Based on the formalism described in Sec. \ref{sec:resoconv}, the power spectrum of the ALP signal for frequencies $\nu_a$ and $\nu_b$ in Rayleigh-Jeans Kelvin ($\rm{K}_{RJ}$) will vary as: 
\begin{equation}
\label{eq:scale}
(C_{\ell}^{ab})_{\rm{ax}} \propto \nu_a \nu_b \,  g_{a\gamma}^{4} \, .  
\end{equation}
These power spectra, evaluated at the standard coupling strength, form the basis for the subsequent statistical analysis aimed at constraining the photon-ALP coupling and the ratios of the signals at microwave and radio frequencies from mock data. This dependence on frequencies forms the vital component that enables the use of CVC technique to constrain the ratios of the signals at different frequencies.
 

Accurate modeling of the ALP-induced power spectrum relies on a detailed characterization of the astrophysical environments in which the conversion takes place, most notably the properties of galaxy clusters. The expected ALP signal depends sensitively on the cluster redshift, the spatial distribution of the electron density, and the structure and strength of the intracluster magnetic field. In this work, we adopt representative values for these quantities, as summarized in Table \ref{tab:params} in Appendix \ref{sec:profiles}, to construct fiducial models for the ALP-induced distortions.
In a realistic observational setting, these cluster properties can be constrained directly using multi-wavelength data. Radio observations provide key information about the intracluster magnetic fields through synchrotron emission and Faraday rotation measurements \cite{murgia2004magnetic,carilli2002cluster,GOVONI_2004,prestage1988cluster}, while X-ray and Sunyaev–Zel’dovich (SZ) \cite{sarazin1986x,Birkinshaw_1999,sunyaev1980microwave,kaastra2004spatially} measurements probe the thermal electron density distribution. The optical and infrared surveys further provide redshift information and cluster identification \cite{bilicki2016wise,zehavi2011galaxy,castagne2012deep,abdalla2011comparison,zaznobin2021spectroscopic}. By combining these complementary datasets, one can characterize the relevant physical parameters of resolvable clusters with significantly improved accuracy, thereby enabling more robust modeling of the expected ALP signal \cite{Mehta:2024wfo,Mehta:2024sye}.
However, not all clusters contributing to the signal will be individually resolved across all observational channels. For such unresolved systems, the cluster-specific parameters—such as the magnetic field strength, electron density profile, and redshift—cannot be directly inferred. The cumulative contribution from these unresolved objects instead manifests as a diffuse ALP-induced background in the observed signal \citep{Mehta:2024pdz,mondino2024axioninducedpatchyscreeningcosmic}.  

\subsubsection{Simulation of the mock signals}\label{sec:obs_powspec}
The primary CMB realization is generated using \texttt{CAMB} \cite{2011ascl.soft02026L}, while the galactic foreground components, specifically thermal dust, free-free and synchrotron emission, are modeled using the ``d-3'', ``f-1'' and ``s-3'' templates from \texttt{PySM} \cite{Thorne_2017} for all the 8 frequency channels (3 bands from SO and 5 bands from SKA). 
The synchrotron emission  model allows for a curved frequency spectrum rather than a simple power law. The free–free emission is assumed to be unpolarized and is modeled using degree-scale smoothed emission-measure and effective electron-temperature templates obtained from the Commander analysis of Planck data \cite{adam2016planck}.
The thermal dust emission model accounts for spatial variations in the dust spectral index .
 These components are combined with the simulated ALP-induced signal at these frequencies corresponding to a given ALP mass ($m_a$) and a fixed fiducial coupling strength, denoted as ${g_{a\gamma}^{\mathrm{true}}}$. The resulting sky maps are then convolved with the instrumental beam and augmented with the appropriate instrumental noise for the corresponding frequency channels. The noise and beam resolutions for SO and SKA are summarized in Table~\ref{tab:noise}.  { At the map-level, the observed fluctuations at a sky-location $\hat{\mathbf{n}}$ are given as the beam-convolved ($B(\hat n, \hat n')$) temperature fluctuations of the true underlying field ($\Delta T^a$), with added instrumental noise ($N^a$):
\begin{equation}
\Delta T_{\mathrm{obs}}^a(\hat{\mathbf{n}})
=
\int d\hat{\mathbf{n}}'\,
B(\hat{\mathbf{n}},\hat{\mathbf{n}}')\,
\Delta T^a(\hat{\mathbf{n}}')
+
N^a(\hat{\mathbf{n}}) \, .
\end{equation}

The power spectrum $C_{\ell}$ is obtained as the two-point correlation of the spherical harmonic coefficients, obtained by expanding the temperature fluctuations in the observed sky, and is explained in Appendix \ref{sec:estimator}.}
Using Eq.~\eqref{eq:modecouple}, where $\mathrm{C_{\ell}'}$ includes contributions from the primary CMB, astrophysical foregrounds, and the ALP-induced signal, we estimate the mock-data power spectrum for the masked map using the \texttt{anafast} routine in \texttt{HEALPy}. We obtain 8 auto-power spectra and 15 cross-power spectra using this method for 3 SO and 5 SKA frequency bands.
Finally, we deconvolve the beam-smoothed power spectra by the beam resolution for the corresponding frequency channels, i.e., 

\begin{equation}
    B_{\ell}^{a}B_{\ell}^{b}C_{\ell}^{ab} + N_{\ell}^{ab} \rightarrow  C_{\ell}^{ab} + (B_{\ell}^{a}B_{\ell}^{b})^{-1}N_{\ell}^{ab} \equiv (C_{\ell}^{ab})_{\rm{data}} \, .
\end{equation}
The net power spectra so obtained is the mock-data power spectrum for the fields $a$ and $b$, denoted as $(C_{\ell}^{ab})_{\rm{data}}$. 

\subsubsection{Fiducial map power spectrum}
\label{sec:fiducial}
We use the power spectrum estimator formalism (explained in Appendix \ref{sec:estimator}) to estimate the contribution of the ALP power spectrum to the mock data power spectrum. This states that:
\begin{equation}
\hat{C}_\ell^{\rm{ax}}
= B_\ell^{-2} \frac{1}{2\ell+1}\sum_{m=-\ell}^{\ell}|a_{\ell m}^{\rm{data}}|^2 - [B_\ell^{-2} N_\ell
+ \sum_{j\neq \rm{ax}} C_\ell^{j}] \, ,
\label{eq:estimator}
\end{equation}
where the first contribution is the mock data power spectrum obtained in Sec. \ref{sec:obs_powspec}. 
The remaining contribution to the observed power spectrum arises from all components other than the ALP-induced signal. We define the ensemble-averaged power spectrum obtained from multiple realizations of these non-ALP components as the fiducial power spectrum. This fiducial spectrum corresponds to the null hypothesis in which no ALP-induced distortion is present and therefore serves as the baseline against which any additional signal is assessed.

 The presence of new physics (here in the form of the ALP signal) can only be inferred through excess fluctuations relative to the expected covariance of the fiducial spectrum. In other words, any statistically significant deviation beyond the variance predicted by the fiducial model can be attributed to an additional contribution such as ALP-induced distortions.

The fiducial power spectrum is constructed by combining the beam-deconvolved contributions from all relevant astrophysical and instrumental components, including the primary CMB and foreground emissions. Explicitly, it is given by the mean of the sum of the power spectra of these components over multiple realizations, providing the reference against which the ALP signal is tested. {Since the power spectra estimation (see Eq. \eqref{eq:covar}) is limited by the cosmic variance, due to observation of a limited number of spherical harmonic modes, multiple realizations are required to account for the variation in estimated power spectrum.} We make 1000 masked fiducial map realizations to obtain the fiducial power spectra. We do so to obtain all the 8 auto and 15 cross fiducial power spectra for the combination of 3 SO bands and 5 SKA bands.

Using the modeled ALP power spectra, the mock-data power spectra, and the fiducial power spectra, we will be able to make an estimate of the ALP coupling and the 15 signal ratios for the cross radio and microwave frequency bands. In the next section, we apply the CVC technique to obtain constraints on these 16 parameters.  

\subsubsection{Bayesian inference with the CVC technique }
We perform Bayesian inference  using the \texttt{emcee} Markov Chain Monte Carlo (MCMC) sampler \cite{Foreman_Mackey_2013}. for the three SO bands separately, so each analysis gives us constraints on 6 parameters (5 ratio parameters and the ALP coupling). We do this for each redshift bin independently.
This gives us the data-vector for analysis corresponding to the $i^{\rm{th}}$ band at each multipole $\ell$ from SO detector as:

\begin{equation}
\mathbf{x}_\ell^{(i)}
=
\begin{pmatrix}
a_i & b_1 & b_2 & b_3 & b_4 & b_5
\end{pmatrix},
\qquad i = 1,2,3.
\end{equation}
where $a_i$ denotes the modes for the $i$-th frequency band of SO, and $b_j$ ($j=1,\dots,5$) correspond to the modes for the frequency bands of SKA. The SO bands 93, 145 and 225 GHz are indexed as $i =$ 1, 2 and 3 respectively, while the SKA bands 0.3, 0.77, 1.4, 6.7, 12.5 GHz are indexed as  $j =$ 1, 2, 3, 4 and 5, respectively.

{We define the ratio parameters for $j^{\rm{th}}$ SKA band and the $i^{\rm{th}}$ SO band as:
\begin{equation}
    \xi_{ij} = \sqrt{\frac{C_{\ell}^{b_jb_j}}{C_{\ell}^{a_ia_i}}} \, .
\end{equation}}
Using the scaling relation for the ALP spectra (see Eq. \eqref{eq:scale}), the theoretical covariance matrix is obtained as the sum of ALP model spectra and the corresponding noises:  
\begin{equation}
\mathbf{C}_\ell^{i}
=
C_\ell^{0}\, (g_{a\gamma } / g_{0})^{4}
\begin{pmatrix}
1
& \xi_{i1}
& \xi_{i2}
& \xi_{i3}
& \xi_{i4}
& \xi_{i5}
\\[4pt]
\xi_{i1} & \xi_{i1}^2 & 0 & 0 & 0 & 0 \\
\xi_{i2} & 0 & \xi_{i2}^2 & 0 & 0 & 0 \\
\xi_{i3} & 0 & 0 & \xi_{i3}^2 & 0 & 0 \\
\xi_{i4} & 0 & 0 & 0 & \xi_{i4}^2 & 0 \\
\xi_{i5} & 0 & 0 & 0 & 0 & \xi_{i5}^2
\end{pmatrix} + [B_\ell^{-2} N_\ell
+ \sum_{j\neq \rm{ax}} C_\ell^{j}]_{i} \, ,
\end{equation}
where $C_\ell^{0}$ sets the overall shape of the angular power spectrum  for $g_{a\gamma} = g_0 = 10^{-12} \, \rm{GeV}^{-1}$ and the common amplitude parameter $g_{a\gamma}^4$ multiplies all components.
{Cross-spectra between bands of the same detector are not being considered as the noises for the same detectors may exhibit correlations, hence, we have set them to zero.} The $ [B_\ell^{-2} N_\ell
+ \sum_{j\neq \rm{ax}} C_\ell^{j}]_{i}$ term refers to the fiducial power spectra matrices with auto and cross spectra components for the $i^{\rm{th}}$ SO band, without the ALP signal.

Assuming Gaussian-distributed spherical-harmonic coefficients and statistical isotropy,
the empirical covariance matrix $\hat{\mathbf{C}}_\ell$ follows a Wishart distribution \cite{hamimeche2008likelihood,planck2016planck}.
The corresponding log-likelihood is \cite{verde2003first,bond1998estimating,tegmark1997measure}
\begin{equation}
-2 \ln \mathcal{L}
=
\sum_{\ell}
(2\ell + 1)
\left[
\mathrm{Tr}\!\left(
\hat{\mathbf{C}}_\ell \mathbf{C}_\ell^{-1}
\right)
+ \ln \det \mathbf{C}_\ell
\right] \, .
\end{equation}
Here $\hat{\mathbf{C}}_\ell$ is the mock-data matrix with the auto and cross-spectra for the $i^{\rm{th}}$ SO band. 
 
The posterior probability distribution for the coupling constant is then obtained using Bayes’ theorem \cite{trotta2017bayesian,heavens2010statisticaltechniquescosmology},
\begin{equation}
\label{eq:bayes}
\mathrm{P(\theta | \mathrm{Data}) \propto \mathcal{L}(\mathrm{Data} | \theta)\, \pi(\theta)} \, ,
\end{equation}
where $\theta$ denotes the set of model parameters and $\pi(\theta)$ represents the prior. We adopt a flat (uninformative) prior on the photon-ALP coupling constant in the range 
\[
10^{-14} \le g_{a\gamma} \le 10^{-11} \ \mathrm{GeV^{-1}} .
\]
For the signal ratios $\xi_{ij}$, we use the flat prior in the range 
\[
0 \le \xi_{ij} \le 1 .
\] 
Limits on the coupling strength are obtained by confronting the appropriately rescaled theoretical power spectrum with the observed spectrum, while fully accounting for the covariance of the power spectrum estimator. Consequently, the strength and robustness of the resulting constraints are largely dictated by the covariance properties of the data, which capture the combined effects of instrumental noise and cosmic variance. In the next section, we look at the results obtained from such an analysis, using only the auto-power spectra, as well as a combination of both the auto and cross-power spectra.

\section{Results}
\label{sec:results}
We perform the Bayesian parameter estimation using \texttt{emcee} \cite{Foreman_Mackey_2013} for the 10 redshift bins with $\Delta z = 0.1$ up (the first bin has $\Delta z = 0.05$) to a redshift of $z = 1$ separately, using the injected value of ALP coupling in mock data as $g_{a\gamma} = 3 \times 10^{-12} \, \rm{GeV}^{-1}$, well below the bounds from CAST at $g_{a\gamma} < 6.6 \times 10^{-11} \, \rm{GeV}^{-1}$ \cite{2017}. Furthermore, the analysis is done separately for each of the three SO bands for each redshift bin. For each $i^{\rm{th}}$ band, we obtain constraints on the parameters $\xi_{i1},\xi_{i2},\xi_{i3},\xi_{i4},\xi_{i5}$ and $g_{a\gamma}$. We use 15 walkers for each run which infers these 6 parameters with a chain size of 25000. We discard the burn-in part and perform thinning for every 20 steps.

For galaxy clusters grouped within a redshift interval of width $\Delta z = 0.1$, the projected angular sizes on the sky are broadly similar. We therefore impose a lower multipole cutoff that corresponds to the largest angular extent of clusters within each redshift bin. This choice mitigates spurious correlations between different clusters arising from large-scale modes, i.e., low multipoles. On small angular scales, the analysis is limited by the instrumental resolution; accordingly, we set the upper multipole cutoff by the beam sizes of the SO frequency channels.

\begin{figure}
    \centering    \includegraphics[width=13cm, height=13cm]{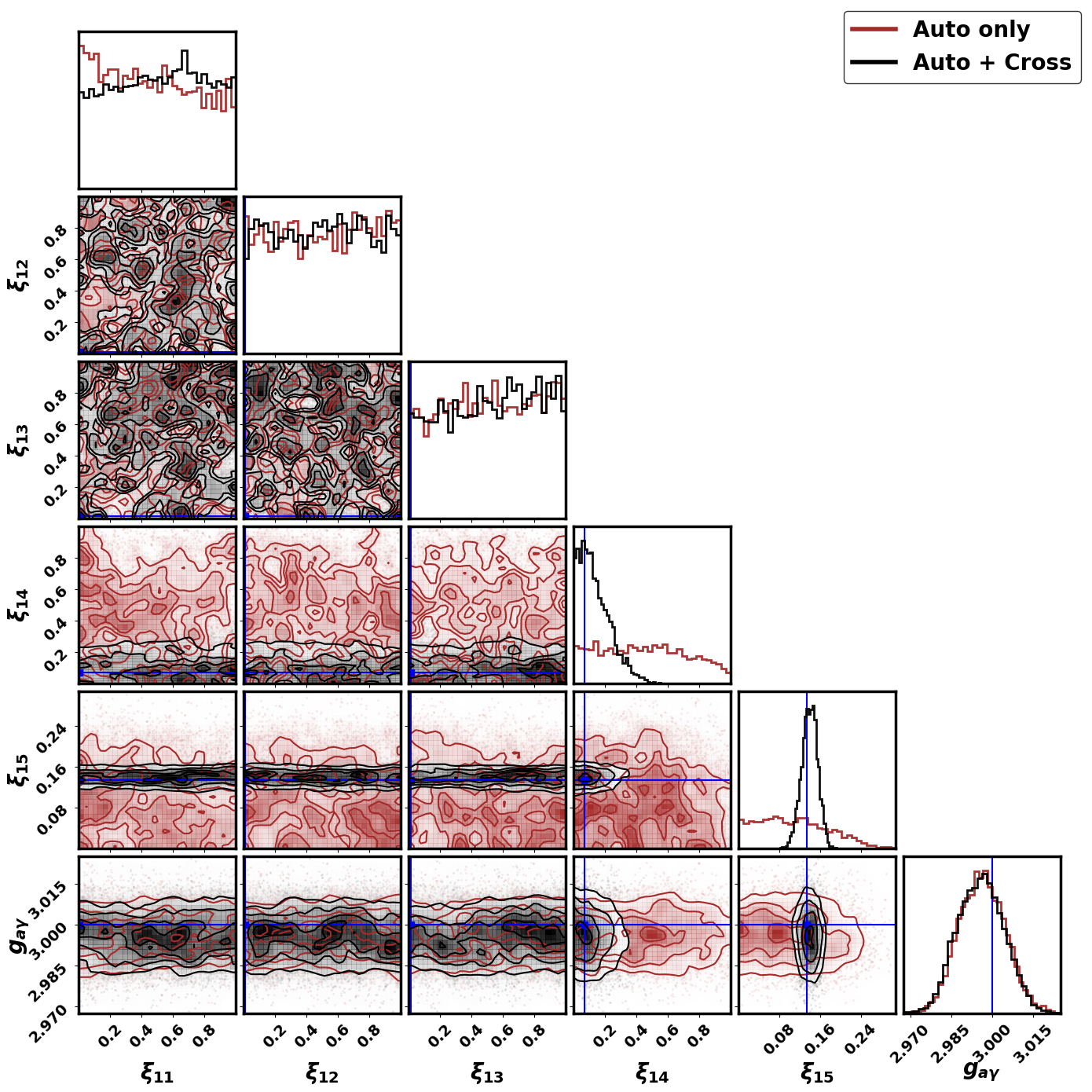}
    \caption{This figure shows the constraints obtained for the 93 GHz band (indexed $i = 1$) for both the auto-only and CVC analyses. Cosmic variance cancellation is apparent for the cases of $\xi_{14}$ and $\xi_{15}$. The injected ALP coupling constant value in mock-data is $g_{a\gamma} = 3 \times 10^{-12} \,  \rm{GeV}^{-1}$.}
    \label{fig:corn1}
\end{figure}

\begin{figure}
    \centering    \includegraphics[width=13cm, height=13cm]{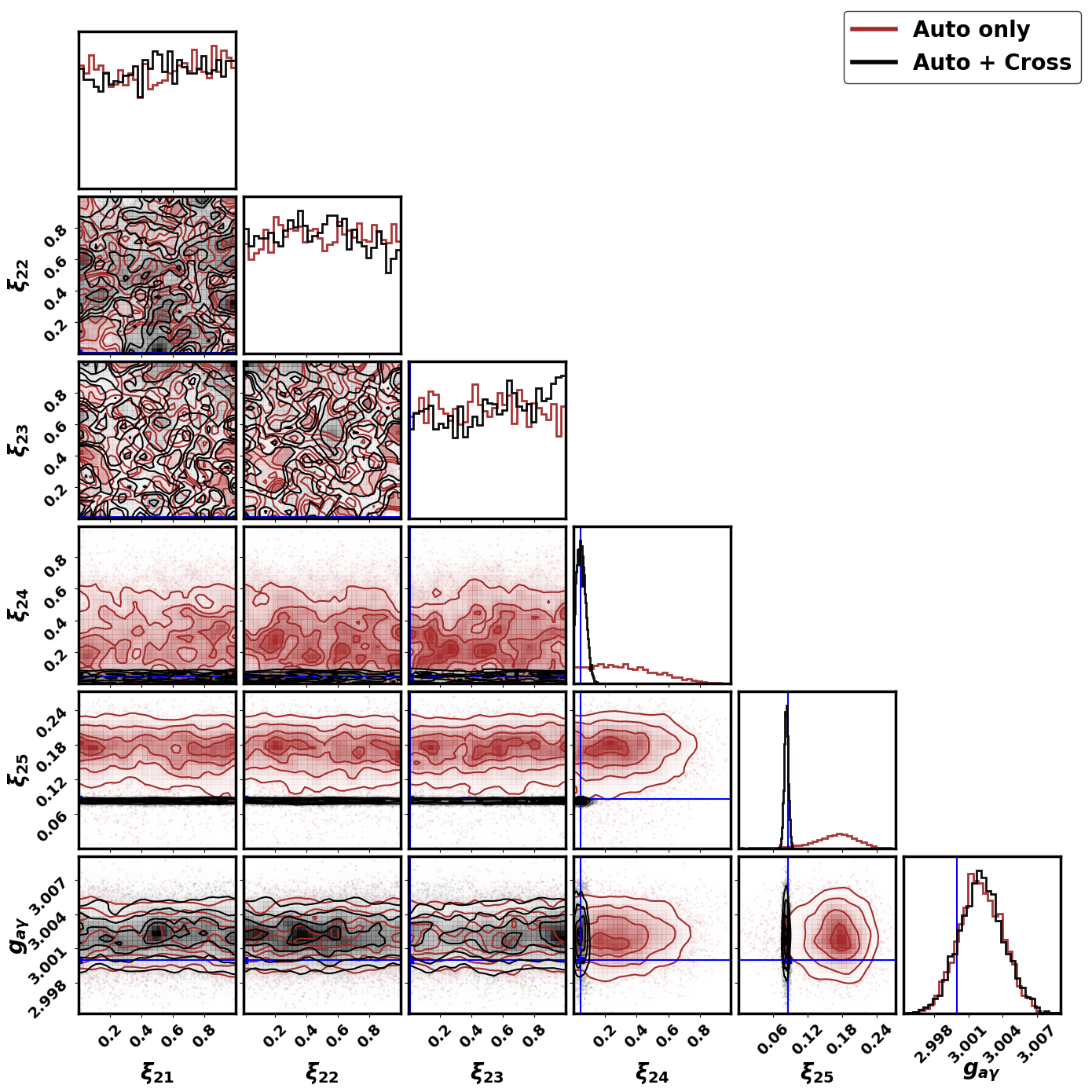}
    \caption{This figure shows the constraints obtained for the 145 GHz band (indexed $i = 2$) for both the auto-only and CVC analyses. Cosmic variance cancellation is apparent for the cases of $\xi_{24}$ and $\xi_{25}$. The injected ALP coupling constant value in mock-data is $g_{a\gamma} = 3 \times 10^{-12} \,  \rm{GeV}^{-1}$.}
    \label{fig:corn2}
\end{figure}

\begin{figure}
    \centering    \includegraphics[width=13cm, height=13cm]{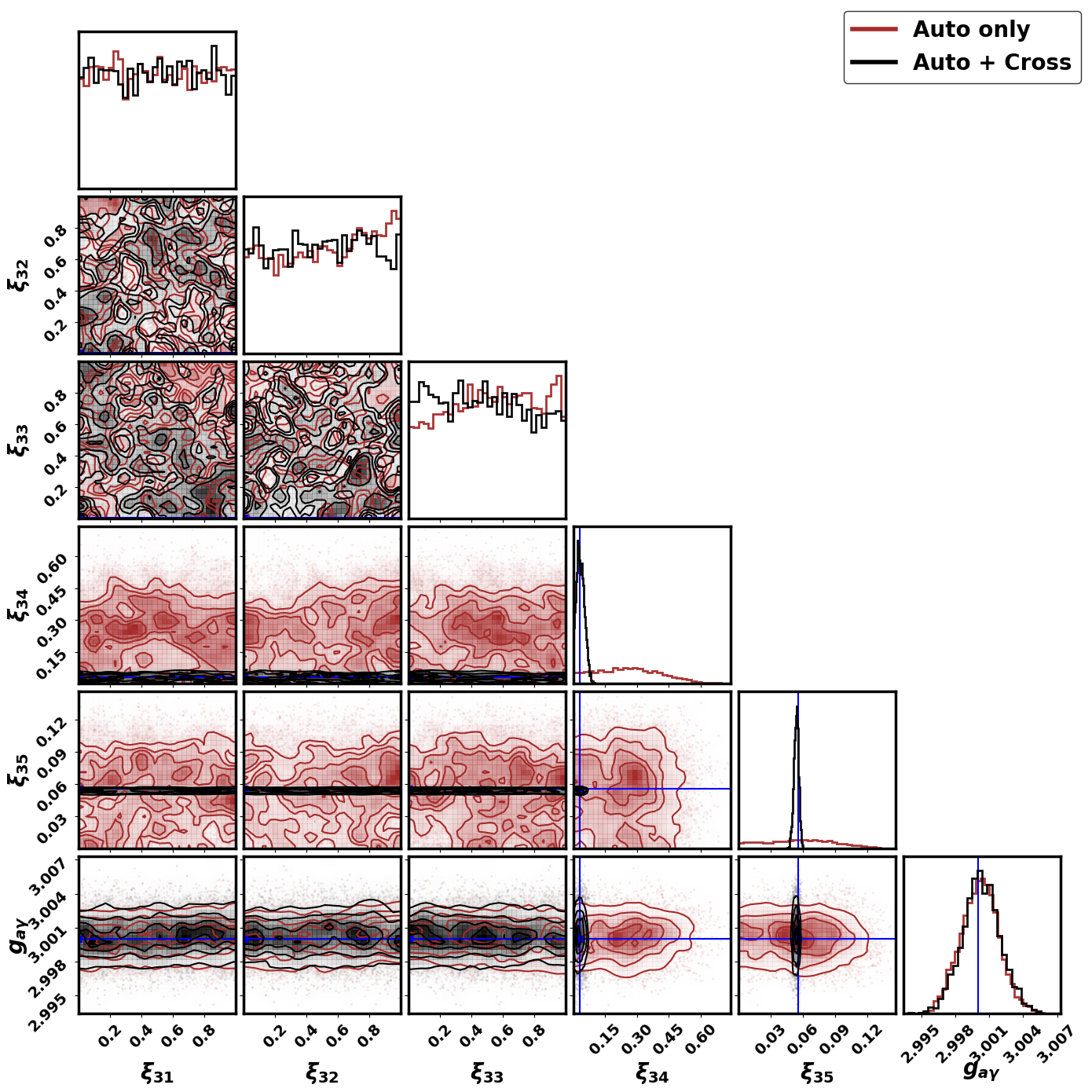}
    \caption{This figure shows the constraints obtained for the 225 GHz band (indexed $i = 3$) for both the auto-only and CVC analysis. Cosmic variance cancellation is apparent for the cases of $\xi_{34}$ and $\xi_{35}$. The injected ALP coupling constant value in mock-data is $g_{a\gamma} = 3 \times 10^{-12} \,  \rm{GeV}^{-1}$.}
    \label{fig:corn3}
\end{figure}
We show the corner plots of the inferences for one of the redshift bins at mean redshift $z = 0.45$ for the three SO bands in Figs. \ref{fig:corn1}, \ref{fig:corn2} and \ref{fig:corn3}, using both the auto-only and CVC combinations. The CVC technique clearly plays a role in the inferences on the ratio parameters related to the SKA bands 6.7 and 12.5 GHz (indexed as $j = 4$ and $j = 5$).
This is mainly a result of the signal being higher in those bands as $\Delta T^{\rm{ax}} \propto \nu$, as well as due to the weakening of the synchrotron  foreground contamination. The constraints are also better for the ratio parameters related to the 225 GHz band of SO (indexed as $i = 3$) as the signal strength is highest in this band and this shows up in the cross spectrum. Observationally, this channel will provide the best constraints on the ratio parameters and synchrotron emission is very weak in the channel and the dust emission is very weak at radio frequencies. {The primary CMB realization is generated using \texttt{CAMB} \cite{2011ascl.soft02026L}, while the galactic foreground components, specifically thermal dust, free-free and synchrotron emission, are modeled using the ``d-3'', ``f-1'' and ``s-3'' templates from \texttt{PySM} \cite{Thorne_2017} for all the 8 frequency channels (3 bands from SO and 5 bands from SKA). }

{{The ALP coupling is primarily inferred from CMB observations, as the corresponding radio signal is expected to be subdominant relative to foreground contamination. Since residual foregrounds can bias these estimates, a joint analysis of the photon–ALP coupling and the signal ratios $\xi_{ij}$ would provide a more robust consistency check, helping to identify and mitigate potential spurious systematic effects that might arise in CMB-only analyses.} This is because biased constraints on the ALP coupling will bias the signal-ratio inferences as well. Using multiple bands from various detectors, the CVC technique will serve as a universal probe of the ALP signal.  }


\begin{figure}
    \centering    \includegraphics[width=13cm, height=8cm]{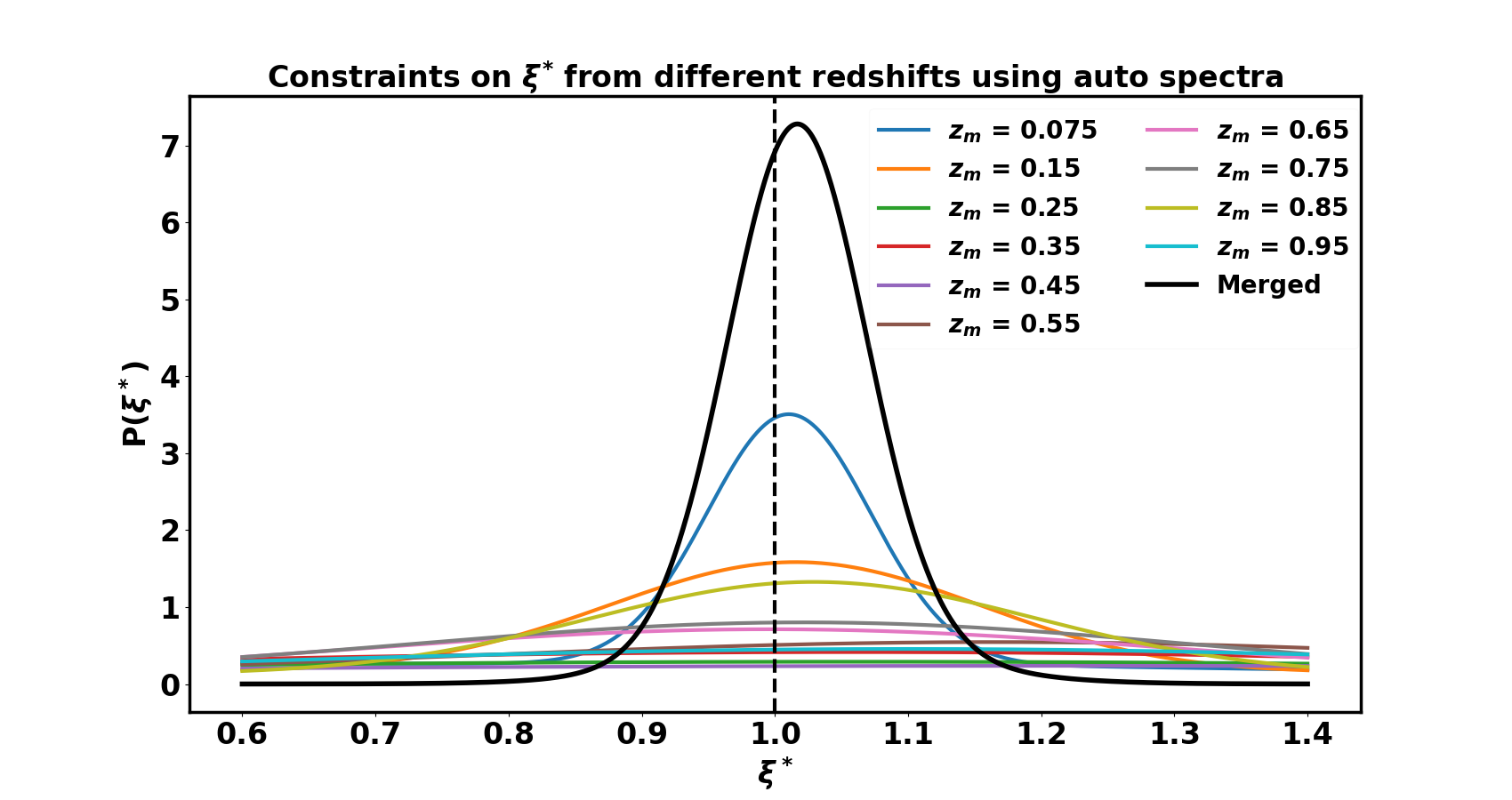}
    \caption{This figure shows the constraints obtained on the normalized parameter $\xi^*$ with the auto-only analysis for various redshift bins. The merged posterior is obtained by multiplying the respective probabilities, followed by subsequent normalization.}
    \label{fig:kdeauto}
\end{figure}

\begin{figure}
    \centering    \includegraphics[width=13cm, height=8cm]{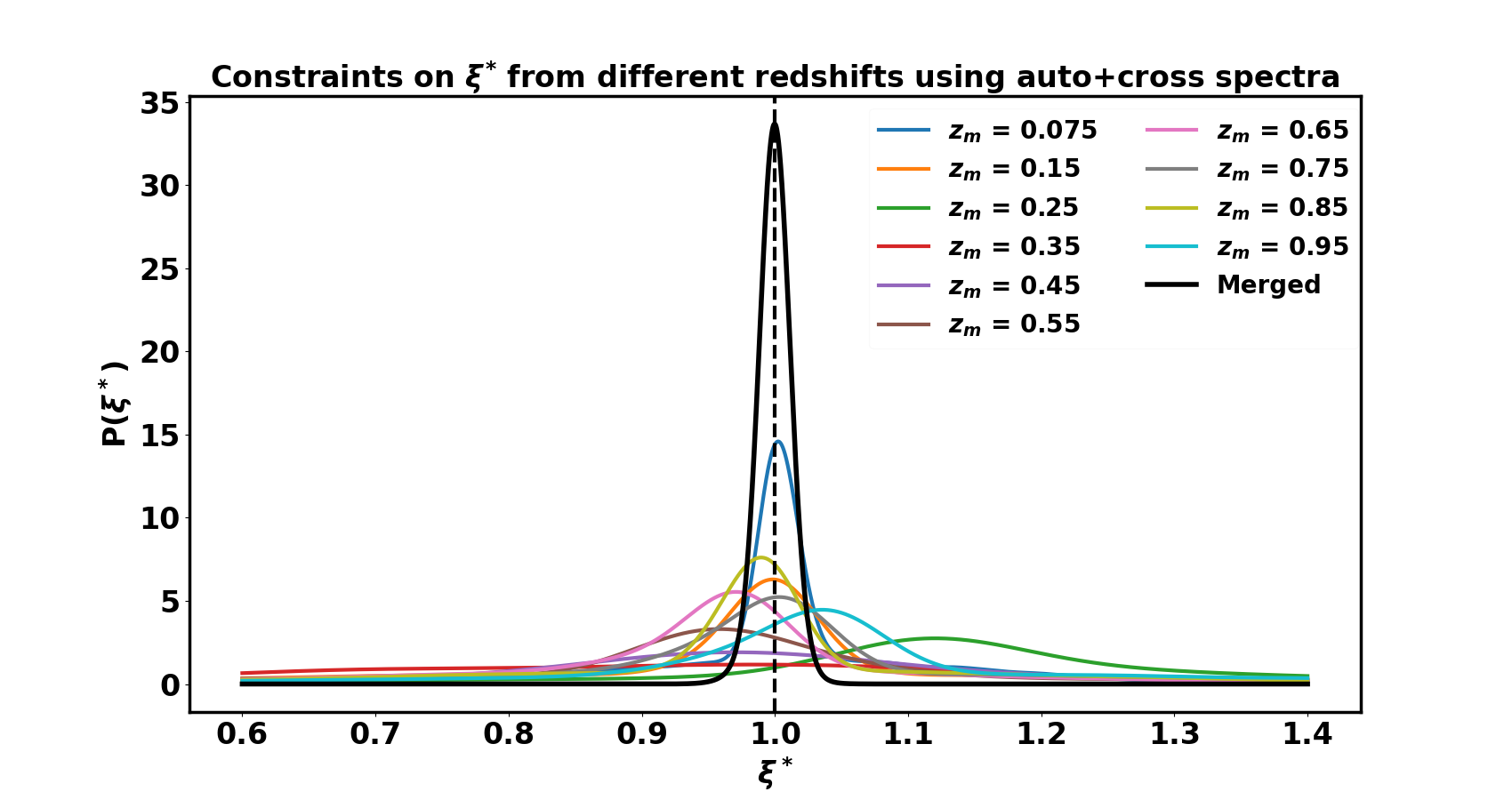}
    \caption{This figure shows the constraints obtained on the normalized parameter $\xi^*$ with the CVC analysis for various redshift bins. The merged posterior is obtained by multiplying the respective probabilities, followed by subsequent normalization. }
    \label{fig:kdecross}
\end{figure}

\begin{table}[htbp]
  \small
  \centering
  \caption{\textbf{Standard deviations on $\xi^*$ from various redshift bins for both the auto-only and CVC combinations.} }
  \label{tab:kdeall}
  \subfloat{
    \begin{tabular}{|c|c|c|}
        \hline
         Redshift bins & $\sigma(\xi^*)^{\rm{auto}}$ & $\sigma(\xi^*)^{\rm{CVC}}$ \\
        \hline
        \hline
        0.05 $\leq z <$ 0.1 & $4.6 \times 10^{-1}$ & $1 \times 10^{-1}$ \\
        \hline
         0.1  $\leq z <$ 0.2 & $9.8 \times 10^{-1}$ & $2.1 \times 10^{-1}$\\
        \hline
         0.2  $\leq z <$ 0.3 & $4.9$ &$4.5 \times 10^{-1}$ \\
        \hline
         0.3  $\leq z <$ 0.4 & $2.9$&$5.5 \times 10^{-1}$\\
        \hline
         0.4  $\leq z <$ 0.5 & $4.5$&$7.4 \times 10^{-1}$ \\
        \hline
         0.5  $\leq z <$ 0.6 & $2.6 $&$4 \times 10^{-1}$ \\
        \hline
         0.6  $\leq z <$ 0.7 & $2.2 $&$1.9 \times 10^{-1}$ \\
        \hline
         0.7  $\leq z <$ 0.8 & $1.7$&$2.3 \times 10^{-1}$ \\
        \hline
         0.8  $\leq z <$ 0.9 & $1.2$ &$1.8 \times 10^{-1}$\\
        \hline
         0.9  $\leq z <$ 1 & $3.2$ &$2.7 \times 10^{-1}$\\
        \hline
         0.05  $\leq z <$ 1 & $5.9 \times 10^{-2}$ &$1.3 \times 10^{-2}$\\
        \hline
    \end{tabular}
    }

\end{table}

The ratio parameters for the low frequency SKA bands (0.3, 0.77 and 1.4 GHz), indexed as $j = 1,2$ and 3, are not very well constrained as the signals are very weak at these low frequencies. With better foreground modeling at radio frequencies, we can use template matching to improve the constraints on all the ratio parameters.  For this analysis, we do not take into account these frequency bands further. {We  analyze the gain in constraints on the ratio parameters from the rest of the bands by combining their information. We normalize  the ratio parameters as
\begin{equation}
    \xi^* = \xi_{ij}\frac{\nu_i}{\nu_j} \, .
\end{equation}
This sets the true value at $\xi^
* = 1$ for the normalized parameter. 
This normalized parameter $\xi^{*}$ will be used to further analyze the gain when combining information from different frequency bands.}

{If the ALP coupling constraints are biased due to foreground contamination, the constraints on $\xi^{*}$  will also be biased. This may shift the constraints from the true value of $\xi^* = 1$.} We combine the histograms so obtained from the 2 ratio parameters (indexed as $j = 4$ and 5) for each redshift bin. We show the posteriors from each redshift bin on $\xi^*$ in Fig. \ref{fig:kdeauto} and Fig. \ref{fig:kdecross} for the cases of auto-only and CVC analyses respectively. The standard deviations on $\xi^*$ for each redshift bin is shown in Table \ref{tab:kdeall} for the cases of auto-only and CVC analyses. The merged posterior from all redshift bins is shown as a bold black line and is obtained by multiplication of probabilities of the inferences from individual bands, and subsequent normalization. The lowest redshift bin $0.05 \leq z < 0.1$ shows the best constraints with standard deviations being $4.6 \times 10^{-1}$ and $1 \times 10^{-1}$ for the auto-only and CVC cases respectively, thus contributing maximally to the combined inference. The combined constraints are dominated by contributions from the lowest redshift bin, as it is this bin that infers the best constraints on $\xi^*$ for both the auto-only and CVC cases.
 This is due to the clusters subtending large angular scales in the sky as compared to the high-redshift clusters.  These larger angular extents amplify the ALP-induced signal fluctuations, leading to improved sensitivity to the ALP--photon coupling.
Moreover, the constraints using the CVC combination is better than those from the auto-only spectra for each redshift bin, as well as when these posteriors are combined.  The CVC technique improves the constraints substantially in all redshift bins, and by an order of magnitude, especially at higher redshifts due to a high number of galaxy clusters. The standard deviation on $\xi^*$ for the combined posterior in auto-only and CVC cases is $5.9 \times 10^{-2}$ and 
$1.3 \times 10^{-2}$ respectively. 
Since it is difficult to obtain smaller error bars, this is a significant improvement.

The sensitivity obtained at different redshifts is primarily driven by a subset of luminous galaxy clusters that generate comparatively large ALP-induced distortions, with the signal amplitude scaling with the assumed coupling strength. The fraction of such bright clusters within a large ensemble, as considered in our analysis, is determined by the underlying distributions of electron density and magnetic field profiles. Although these cluster properties are not tightly constrained observationally, our results are not significantly affected when the profile parameters are drawn from uniform random distributions, since deviations from the mean tend to average over a sufficiently large cluster sample. In practice, however, the resulting constraints are governed by the probability of encountering bright clusters within the population, which itself depends on the distribution of magnetic field strengths and electron densities across clusters as a function of redshift.


\begin{figure}
    \centering    \includegraphics[width=13cm, height=8cm]{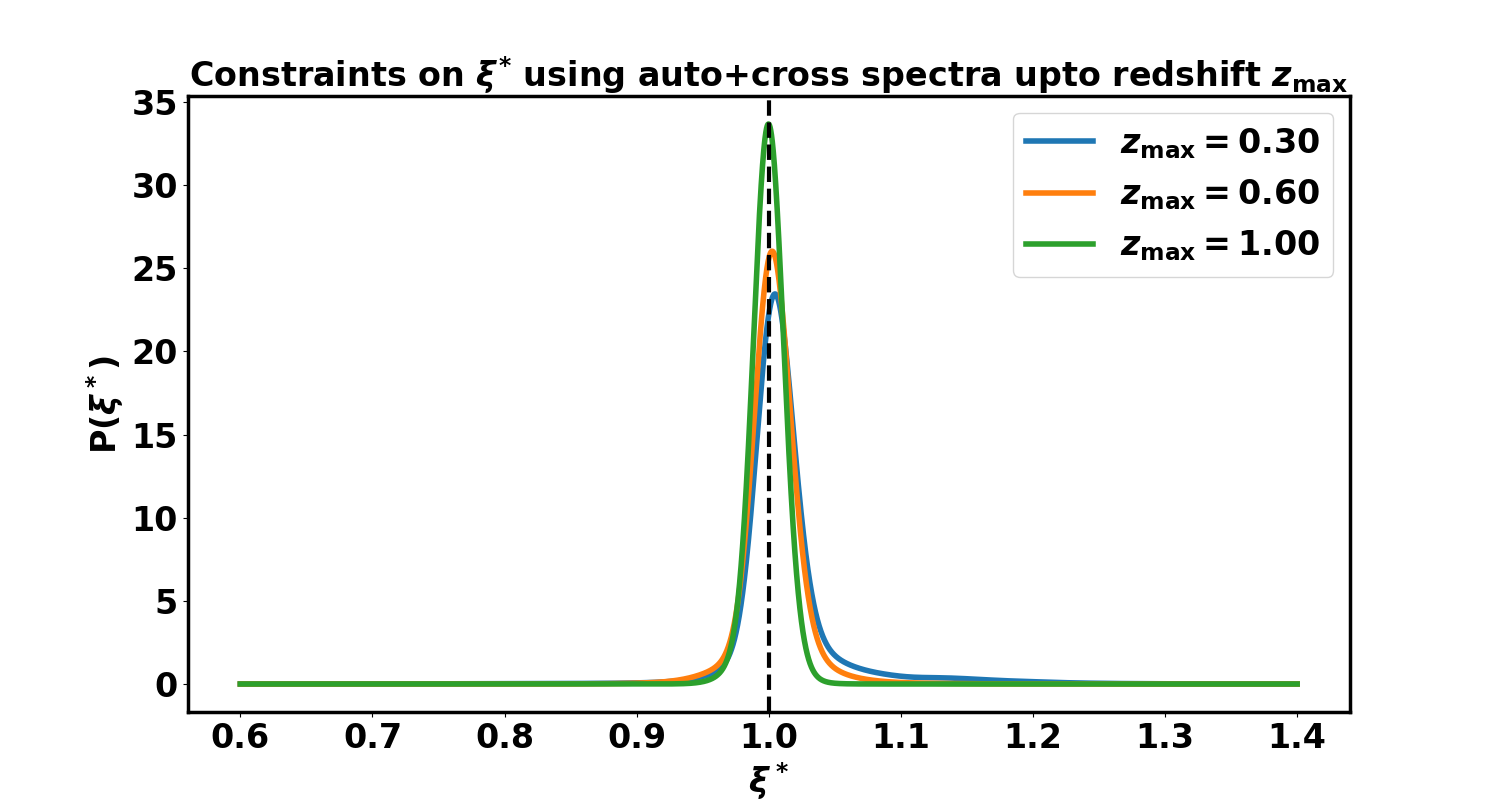}
    \caption{This figure shows the constraints obtained on the normalized parameter $\xi^*$ with the CVC analysis up to the maximum redshift $z_{\rm{max}}$.  The constraints improve for higher $z_{\rm{max}}$. }
    \label{fig:maxcross}
\end{figure}

\begin{table}[htbp]
  \small
  \centering
  \caption{\textbf{Standard deviations on $\xi^*$ upto a redshift $z_{\rm{max}}$ for both the auto-only and CVC combinations.} }
  \label{tab:kdemax}
  \subfloat{
    \begin{tabular}{|c|c|c|}
        \hline
         Maximum redshift & $\sigma(\xi^*)^{\rm{auto}}$ & $\sigma(\xi^*)^{\rm{CVC}}$ \\
        \hline
        \hline
        $ z <$ 0.3 & $8.3 \times 10^{-2}$ & $3.9 \times 10^{-2}$ \\
        \hline
          $ z <$ 0.6  & $7.7 \times 10^{-2}$ & $2.1 \times 10^{-2}$\\
        \hline
          $ z <$ 1  & $5.9 \times 10^{-2}$ & $1.3 \times 10^{-2}$ \\
        \hline
    \end{tabular}
    }

\end{table}

In Fig. \ref{fig:maxcross}, we show the merged posteriors on $\xi^*$ for all redshifts up to a certain $z_{\rm{max}}$ for the  CVC cases. The constraints improve with higher $z_{\rm{max}}$ as more clusters contribute to increased improvements. The standard deviations for both the auto-only and CVC cases up to various $z_{\rm{ max}}$ values is shown in Table \ref{tab:kdemax}
Also, using CVC up to the lowest $z_{\rm{max}} = 0.3$ provides far better constraints than the auto-only constraints up to $z_{\rm{max}} = 1$. This shows that using CVC much more information can be extracted, which substantially improves the constraints.

\begin{figure}
    \centering    \includegraphics[width=13cm, height=8cm]{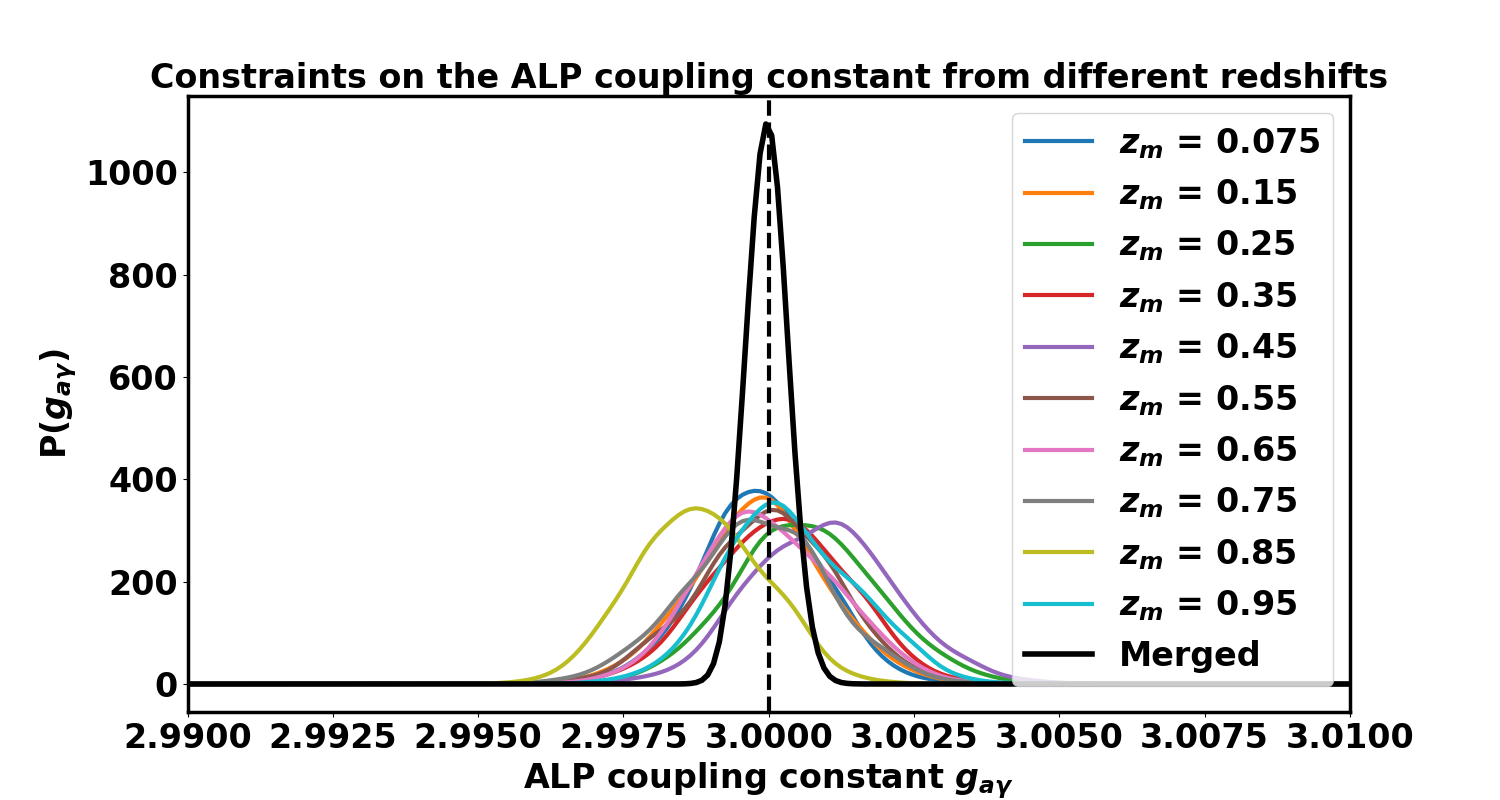}
    \caption{This figure shows the constraints obtained on the photon-ALP coupling constant $g_{a\gamma}$ with the CVC analysis for various redshift bins. The merged posterior is obtained by multiplying the respective probabilities, followed by subsequent normalization. }
    \label{fig:gacons}
\end{figure}
{In Fig. \ref{fig:gacons}, we show the constraints obtained on the photon-ALP coupling for all the redshift bins. The precision is limited by the cosmic variance and is mainly contributed to by the CMB observations, as the radio signal is very weak when compared to the foreground contamination. The constraints on the ALP coupling although do not improve when compared to the auto-only analysis, these are important as any bias on the ALP coupling will affect the inference on the ratio parameters and hence, the normalized ratio $\xi^*$.}
 {The constraints also depend sensitively on the ALP mass. The lower-mass ALPs generate larger distortion signatures, resulting in stronger constraints compared to heavier ALPs. The CVC technique thus performs better for lower mass ALPs and can be used to study their signatures. Since these ALPs are generated in outer regions of clusters (see Equ. \eqref{eq:resonance mass}) and may suffer from higher uncertainty in cluster profiles, the CVC technique serves as an important investigator {of any potential contaminants or systematics} using the spectral signature of the ALP distortion signal in different bands.
}

{Using multiple frequency bands is crucial for separating a potential ALP-induced signal from astrophysical foregrounds, since the latter typically exhibit distinct spectral behavior while the ALP contribution follows a different, model-dependent scaling. In addition, cosmic variance cancellation exploits the fact that different observables tracing the same underlying large-scale structure share the same realization of primordial fluctuations. By correlating datasets such as CMB and radio maps—each with a different response to ALP physics but overlapping modes—one can reduce the sample variance associated with those modes and more effectively isolate the subdominant ALP signal. Importantly, the inclusion of radio channels also provides a valuable consistency check on CMB-based inferences. Since the ALP signal should manifest in a correlated manner across both datasets, agreement between them strengthens the interpretation, while discrepancies can help flag residual foregrounds or systematic effects in CMB-only analyses. This formalism improves both the sensitivity and the reliability of constraints on the photon–ALP coupling. 
}

{The constraints can be improved by performing a multi-band analysis using information from X-rays (say, using eROSITA \cite{merloni2012erosita,predehl2021erosita,bulbul2024srgerosita}), infrared (say, using WISE \cite{wen2018catalogue}) and other bands. Since the  photon-ALP conversion in these bands may exhibit higher conversion probabilities, the CVC technique will be able to infer much more information and would provide a more robust consistency check, helping to identify and mitigate potential spurious systematic effects that might arise in CMB-only analyses. This technique can thus be used to study various sorts of signatures from different dark matter candidates in different frequency regimes. The projected sensitivity can also be substantially improved with next-generation experiments such as CMB-HD, which provide higher angular resolution and lower instrumental noise.} {Add more on the CVC for other bands and how it can be better.}

At higher redshifts ($z > 1$), galaxy clusters may not be individually resolved across multiple electromagnetic bands with current experiments, limiting our knowledge of their electron density and magnetic field profiles. The ALP-induced distortions from these unresolved clusters therefore combine to form a diffuse background signal, which correlates with the large scale structure \cite{Mehta:2024pdz,mondino2024axioninducedpatchyscreeningcosmic}. The ALP signal from these high-redshift clusters will also show up in radio frequency channels and CVC can also be used to extract more information from them.

{In this work, we have considered the distortion signal at the intensity level (see Eq. \eqref{eq:polint}), which requires the transverse magnetic field magnitude information to probe the ALP signal. Using radio synchrotron polarization measurements, the transverse magnetic field profile in terms of both magnitude as well as direction can be inferred \cite{Mehta:2024sye}, which will provide information on the ALP distortion signal at the level of individual $Q$ and $U$ maps. Using this information, a rotated stacking of ALP distortion polarization signals could further improve the possibility of detecting the ALP signal. }
\section{Conclusion}\label{sec:conclusion}
Observations of a single tracer in the sky suffer from cosmic variance limited inferences \cite{Dodelson:2003ft,white1993cosmic,kamionkowski1997getting,colombi2000experimental,szapudi1995cosmic,driver2010quantifying,somerville2004cosmic}.
 As observational sensitivities continue to improve, cosmic variance increasingly emerges as a fundamental limitation rather than an instrumental one. The ability to mitigate the impact of cosmic variance plays a central role in extending the scientific reach of present and future cosmological analyses \cite{baleato2023model,sibthorpe2012extragalactic,seljak2009extracting,foreman2019cosmic,schmittfull2018parameter}. 

The impact of CMB photon-ALP resonant conversion on the CMB power spectrum has been studied in our previous works \cite{Mehta:2024pdz,Mehta:2024wfo,Mehta:2024sye,Mehta:2025slu}. We will be able to obtain ALP coupling bounds at $g_{a\gamma} < 5.2 \times 10^{-12} \, \mathrm{GeV}^{-1}$ and $g_{a\gamma} < 3.6 \times 10^{-12} \, \mathrm{GeV}^{-1}$ with SO and CMB-S4 respectively for ALP mass $m_a = 10^{-13}$ eV \cite{Mehta:2024wfo}. But these coupling constraints suffer from cosmic variance limited inference, as well as they may be biased by residual foreground contamination, giving false estimates on the ALP coupling. 

If ALPs exist in nature, such a conversion will show signatures at radio frequencies as well. Since this low frequency regime is highly dominated by foregrounds, the auto-power spectrum does not hold much information. But the cross-power spectra of radio and microwave signals can be used to extract additional information using CVC \cite{baleato2023model,sibthorpe2012extragalactic,seljak2009extracting,foreman2019cosmic,schmittfull2018parameter}, as the cross instrumental noise is negligible and the correlation between the contaminating foregrounds at these frequencies is weak. This does not significantly improve the constraints on the ALP coupling, but drastically does so for the ratio of the ALP signal in the two regimes.
Not only does the CVC technique open a new window to detect ALPs using a proxy instead of the ALP coupling, but
can also confirm the universal behaviour of ALPs using the spectral variation of the ALP distortion signal in the two regimes.

In our work, we analyze the improvement in the constraints on the ratio of the signals in SO and SKA bands, when using the CVC technique. We perform our analysis for SO observed clusters up to a redshift of $z = 1$, in bins of $\Delta z = 0.1$. The improvement is substantial for the ratios related to the high-frequency radio bands (6.7 and 12.5 GHz), as the ALP distortion signal weakens significantly at lower frequencies ($\Delta T^{\rm{ax}} \propto \nu$). The CVC technique improves the constraints substantially in all redshift bins, and by an order of magnitude, especially at higher redshifts due to a higher number of galaxy clusters. The standard deviation on the normalized parameter $\xi^*$ is obtained as $5.9 \times 10^{-2}$ and $1.3 \times 10^{-2}$ in the auto-only and CVC cases, being driven by the lowest redshift contributions due to the clusters being highly resolved at low redshifts. Since it is difficult to obtain smaller error bars, this is a significant improvement. This technique would be very beneficial in being able to confirm the detection of any ALP distortion signal and provide a consistency check of ALP signal inferences, using the correlation between multi-frequency tracers.

In our analysis, we have considered clusters with smooth electron density and magnetic field profiles. In reality, the CMB photon-ALP conversion will be impacted by the turbulence in these clusters, affecting the ALP distortion signal \cite{Mehta:2025qfr}. This will cause multiple resonant locations, along with depolarization of the signal due to misaligning magnetic field domains along the line of sight.
This will affect the power spectrum of the ALP signal, but the CVC technique can be used across different frequency bands, as it makes use of the spectral information of the signal. Additionally, the breakdown of the Landau-Zener approximation for such  profiles can change the correlation between the ALP distortion spectra at different frequencies \cite{zener1932non,carenza2023applicability,marsh2022fourier}.

The CVC technique is particularly useful in detecting low mass ALPs, which will generate stronger signals with larger disk sizes as compared to high mass ALPs \cite{Mehta:2024wfo}. Moreover,
the CVC technique can be applied using other frequencies, apart from radio and microwave as well, such as infrared surveys, such as WISE \cite{wen2018catalogue} or X-ray surveys, such as eROSITA \cite{merloni2012erosita,predehl2021erosita,bulbul2024srgerosita}.  This can further suppress the cosmic variance and improve inferences. Thus, using a multi-band inference, we can not only detect ALPs, but probe their universal nature as well and check for any potential contaminants or systematic effects using the spectral behaviour of the ALP signal. Moreover, SKA possess beams of about a few arcseconds \cite{braun2019anticipated}. This would allow future CMB experiments such as CMB-HD \cite{sehgal2019cmbhd} to probe the ALP signal at much higher resolutions along with SKA using CVC. {Also, the CMB EB cross spectrum measurements can be used to probe the ALP signal as the rotation of the polarization plane due to resonant conversion will generate parity-violating modes \cite{ferreira2024axionic,jain2021cmb,jain2022searching,zagatti2024planck,greco2022cosmic,grain2012cmb}. The ALP distortion signal will contribute to the cross spectrum at small scales corresponding to the cluster regions. With the upcoming CMB experiments, this will be an area that can be explored in the future.}

 {In this work, we model both the SO and SKA surveys as beam-convolved maps, which provides a useful first-order approximation for forecasting cosmic variance cancellation. However, we note that the SKA operates as an interferometer measuring visibilities with incomplete and anisotropic uv-coverage \cite{Carilli_2004,braun2019anticipated,subrahmanyan2004radio}. This leads to a scale- and direction-dependent transfer function that can suppress or filter certain spatial modes, particularly on large angular scales. In contrast, SO produces maps through a scanning strategy with a different effective filtering \cite{Ade_2019,hanany2012cmb}. Since cosmic variance cancellation relies on overlapping modes between datasets, such differences in mode recovery and transfer functions may reduce the achievable improvement compared to the idealized case considered here. In future, application of this technique on the observed data would require incorporating the interferometric response and imaging pipeline of SKA, which we leave for future work.}
 
 The CVC technique explored in this work highlights how combining multiple probes that trace the same underlying phenomenon or field, can partially circumvent this limitation, enabling access to information that would otherwise remain hidden by cosmic variance. This calls for multi-band observations by exploiting correlated signals across different frequency regimes, enhancing the effective statistical power of existing and upcoming data sets and open new avenues for probing subtle physical effects. In this context, cosmic variance cancellation is not merely a technical refinement, but a critical ingredient in maximizing the scientific return of next-generation cosmological observations.

\appendix
\section*{Appendices}

\section{Cosmic Variance as a Fundamental limit on Single-field observations} \label{sec:estimator}
The anisotropies observed in the Cosmic Microwave Background (CMB) arise from a variety of physical processes acting at different epochs of cosmic evolution. These temperature and polarization fluctuations are encoded on the celestial sphere and can be represented as two-dimensional fields defined over angular coordinates. A natural and convenient way to analyze such fields is through their decomposition into spherical harmonics, which form a complete orthonormal basis on the sphere. Following standard treatments \cite{Dodelson:2003ft,Hu_2002}, the observed sky signal can be expanded as
\begin{equation}
\Delta^{\rm net}(\theta,\phi) = \sum_{\ell=0}^{\infty}\sum_{m=-\ell}^{\ell} a_{\ell m} Y_{\ell m}(\theta,\phi) \, ,
\label{eq:sphdec}
\end{equation}
where the coefficients $a_{\ell m}$ encode the statistical information of the sky fluctuations, and $\ell$ and $m$ denote the angular multipole and its azimuthal index, respectively.
The observed CMB sky is a superposition of several physically distinct contributions. In addition to the primordial CMB signal, it includes foreground emissions, instrumental noise, and any additional astrophysical or exotic components. This can be written schematically as
\begin{equation}
\Delta^{\rm net} = \Delta^{\rm CMB} + \sum_{i \neq {\rm CMB}} \Delta^{i},
\label{eq:pertparts}
\end{equation}
where each term corresponds to a different physical source of anisotropy. The statistical properties of the sky are therefore encoded in the ensemble averages of the spherical harmonic coefficients. Under the assumption of statistical isotropy, the two-point function takes the diagonal form
\begin{equation}
\langle a^{\rm net}_{\ell m} a^{\rm net,*}_{\ell' m'} \rangle = \delta_{\ell \ell'} \delta_{m m'} C_\ell^{\rm net},
\label{eq:cl_alm}
\end{equation}
where $C_\ell^{\rm net}$ denotes the total angular power spectrum.
In practice, the observed coefficients receive contributions from multiple independent components, including the CMB, astrophysical foregrounds, and potential exotic signals such as axion-induced distortions. Assuming these contributions are mutually uncorrelated, the total power spectrum may be written as
\begin{equation}
C_\ell^{\rm net} = \sum_i C_\ell^{i} \, ,
\label{eq:clnet}
\end{equation}
where the index $i$ labels the different components. Observations are further affected by instrumental effects, most notably the finite angular resolution of the telescope and detector noise. These effects modify the observed spherical harmonic coefficients as
\begin{equation}
a_{\ell m}^{\rm obs} = B_\ell \left( a_{\ell m}^{\rm cmb} + a_{\ell m}^{\rm ax} + a_{\ell m}^{\rm fg} \right) + \eta_{\ell m} \, ,
\label{eq:alm_obs}
\end{equation}
where $B_\ell = \exp[-\ell(\ell+1)\theta_{\rm beam}^2/2]$ is the instrumental beam transfer function, and $\eta_{\ell m}$ represents instrumental noise with variance
\begin{equation}
\langle \eta_{\ell m} \eta_{\ell' m'}^{*} \rangle = N_\ell \delta_{\ell\ell'} \delta_{m m'} \, .
\label{eq:Nl}
\end{equation}
Assuming Gaussian statistics, the probability distribution for the harmonic coefficients conditioned on a given power spectrum is
\begin{equation}
P(a_{\ell m}^{i}|C_{\ell}^{i}) = \frac{1}{\sqrt{2\pi C_\ell^i}}
\exp\left(-\frac{|a_{\ell m}^{i}|^2}{2C_\ell^i}\right),
\label{eq:prob_alm_cl}
\end{equation}
and the likelihood for the observed coefficients can be obtained by marginalizing over the individual sky components:
\begin{equation}
P(a_{\ell m}^{\rm obs}|{C_\ell^i}) =
\prod_{m=-\ell}^{\ell}
\int \prod_i da_{\ell m}^i,
P(a_{\ell m}^{\rm obs}|a_{\ell m}^i)
P(a_{\ell m}^i|C_\ell^i) \, .
\label{eq:likeBayes}
\end{equation}
Carrying out this integration yields a Gaussian likelihood of the form
\begin{equation}
\mathcal{L} =
\left[2\pi\left(B_\ell^2 \sum_i C_\ell^i + N_\ell\right)\right]^{-(2\ell+1)/2}
\exp\left[-\frac{1}{2}\sum_{m=-\ell}^{\ell}
\frac{|a_{\ell m}^{\rm obs}|^2}{B_\ell^2\sum_i C_\ell^i + N_\ell}
\right],
\label{eq:problike}
\end{equation}
from which the maximum-likelihood estimator for the power spectrum of a given component $i$ follows as
\begin{equation}
\tilde{C}_\ell^{,i}
= B_\ell^{-2}\left[\frac{1}{2\ell+1}\sum_{m=-\ell}^{\ell}|a_{\ell m}^{\rm obs}|^2 - N_\ell\right]
- \sum_{j\neq i} C_\ell^{j}.
\label{eq:estimator}
\end{equation}
The uncertainty associated with this estimator is quantified by its covariance (Cov), which reflects the finite number of available modes on the sky. For partial-sky coverage, the covariance increases due to mode loss and is given by
\begin{equation}
\sqrt{\mathrm{Cov}(\tilde{C}_\ell)}
= \frac{2}{(2\ell+1) f_{\rm sky}}
\left(\sum_i C_\ell^i + B_\ell^{-2} N_\ell \right) ,
\label{eq:covar}
\end{equation}
where $f_{\rm sky}$ denotes the observed sky fraction. It shows contribution from two terms: the cosmic variance term $\frac{2}{2\ell + 1}C_{\ell}$ which is a consequence of single field realization in the sky. The second term depends on the instrument noise and beam resolution. With upcoming detectors such as SKA and SO, with low noise and high resolution, the contribution of the latter term $B_{\ell}^{-2} N_{\ell}$ drops significantly. The cosmic variance though, places a fundamental limit on the precision of our measurements and cannot be escaped with a single field observation.  

In the context of this work, the ALP-induced contribution to the angular power spectrum is constrained by comparing its predicted imprint against the covariance-dominated uncertainty of the fiducial (non-ALP) sky. This framework enables a statistically consistent assessment of the detectability of axion-induced signatures in the presence of instrumental noise, cosmic variance, and astrophysical foregrounds.

\section{Cluster modeling and ALP-induced distortion}
\label{sec:profiles}
\begin{table}[h!]
\caption{A list of parameters used in estimating the ALPs signal and their significance}
\label{tab:params}
\hspace{-2cm}
\begin{tabular}{|c|c|c|c|}

\hline
\textit{\textbf{Notation}} &
\textit{\textbf{Description}} &
\textit{\textbf{Typical Estimate}} &
\textit{\textbf{Range}} 

\tabularnewline \hline
s & Magnetic field steepness  & 0.5 & $\mathrm{0.5 < s < 2}$ \\
 \hline
 $\mathrm{B_0}$ & Order of magnetic field at 1 Mpc & 3 & $\mathrm{0.01 < B_0 < 0.5}$  \\ 
 \hline
$\mathrm{n_{02}}$ & Inner region electron density order & $\mathrm{10^{-1} \, cm^{-3}}$ & $\mathrm{5\times 10^{-2} \, cm^{-3} < n_{02}  < 1.5 \times 10^{-1} \, cm^{-3}}$  \\
 \hline
$\mathrm{n_{0}}$ & Outer region electron density order & $\mathrm{10^{-3} \, cm^{-3}}$ & $\mathrm{5\times10^{-4}\, cm^{-3} < n_{0}  < 1.5 \times 10^{-3} \, cm^{-3}}$  \\ 
 \hline
$\gamma$ & Transition width parameter & 3 & $ 2 < \gamma < 4$  \\
 \hline
$\alpha$ & Cusp slope parameter & 2 & $1 < \alpha < 3$ \\ 
 \hline
$\beta_1$ & Outer $ \beta$ exponent & 0.64 & $0.5 < \beta_1 < 0.8$  \\ 
\hline
$\beta_2$ & Inner $ \beta$ exponent & 1 & $0.8 < \beta_2 < 1.2$ \\ 
 \hline
$\mathrm{r_{s}}$ & Scaling radius & $\mathrm{1 \, Mpc}$ & $\mathrm{0.5 \, Mpc < r_s < 1.5 \, Mpc}$  \\ 
 \hline
$\mathrm{r_{c1}}$ & Outer core radius & $\mathrm{0.1 \, Mpc}$ & $\mathrm{0.05 \, Mpc < r_{c1} < 0.7 \, Mpc}$  \\ 
 \hline
$\mathrm{r_{c2}}$ & Inner core radius & 
  $\mathrm{0.01 \, Mpc}$ & $\mathrm{0.008 \, Mpc < r_{c2} < 0.05 \, Mpc }$ \\ 
 \hline
$\epsilon$ & Knee slope & 4 & $2 < \epsilon < 5$ \\ 
 \hline
 $\mathrm{Z}$ & Metallicity & 1.12 & $ 0.8 < \mathrm{Z} < 1.5$\\ 
 \hline
$\mathrm{g_{a\gamma}}$ & ALP coupling constant & $ < \mathrm{10^{-11} \, GeV^{-1}}$ & $\mathrm{10^{-14}\, GeV^{-1} < \mathrm{g_{a\gamma}} < 10^{-11} \, GeV^{-1}}$\\ [1ex] 
 \hline
\end{tabular} 
\end{table}

The axion-like particle (ALP)–induced distortion signal is intrinsically linked to the astrophysical properties of galaxy clusters, as both the production and subsequent propagation of ALPs depend sensitively on the local plasma and magnetic field environment. In this work, clusters are modeled as spherically symmetric systems such that all physical quantities depend solely on the radial distance from the cluster center. This approximation captures the dominant large-scale structure of relaxed clusters and provides a controlled framework for computing the expected ALP-induced signal.

We neglect small-scale turbulent fluctuations in the magnetic field and electron density, assuming instead smooth radial profiles. This approximation is justified when the coherence length of the ordered magnetic field exceeds the characteristic scale over which the resonant photon–ALP conversion occurs, or equivalently, when the electron density varies slowly over the conversion region. In this regime, turbulent depolarization effects are subdominant, and the resulting ALP-induced polarization signal remains coherent.

The thermal electron density profile is modeled using a generalized double–$\beta$ profile, which has been shown to successfully describe observed cluster gas distributions \citep{Vikhlinin_2006,mcdonald2013growth}. The profile is given by
\begin{equation}
{
n_e^2(r) = Z\left[
n_0^2
\frac{(r/r_{c1})^{-\alpha}}{(1 + r^2/r_{c1}^2)^{3\beta - \alpha/2}}
\frac{1}{(1 + r^{\gamma}/r_s^{\gamma})^{\epsilon/\gamma}}
+ \frac{n_{02}^2}{(1 + r^2/r_{c2}^2)^{3\beta_2}}
\right],
}
\label{eq:elec_dens}
\end{equation}
where the various parameters control the inner cusp, outer slope, and transition behavior of the gas distribution. This functional form captures key observational features such as a dense central core,  gradual flattening at intermediate radii, and a steep decline in the outskirts of the cluster. The  profile parameters and their ranges are shown in Table \ref{tab:params}.

The magnetic field structure in galaxy clusters remains uncertain, particularly on small spatial scales. For the purposes of this analysis, we adopt a phenomenological power-law parameterization consistent with observational constraints from Faraday rotation and synchrotron studies \cite{bonafede2010galaxy,bohringer2016cosmic,carilli2002cluster}:
\begin{equation}
{B(r) = B_0\, r^{-s}},
\label{eq:mag_prof}
\end{equation}
where $B_0$ denotes the characteristic magnetic field strength and $s$ controls its radial decline. The field orientation is assumed to be randomly oriented at different locations within the cluster, an assumption that captures the statistical isotropy of the magnetic field on scales larger than the coherence length. The  profile parameters and their ranges are shown in Table \ref{tab:params}.
 The electron density and magnetic field profile parameters are varied within ranges chosen to encompass the diversity of cluster environments inferred from X-ray and radio observations of low-redshift clusters.

ALPs produced through resonant photon–ALP conversion satisfy the condition $m_a = m_\gamma$, where the effective photon mass is set by the local plasma frequency and hence, by the electron density. As a consequence, ALPs of different masses are generated at different radii within the cluster. Higher-mass ALPs correspond to higher plasma frequencies and are therefore produced closer to the dense cluster core, while lower-mass ALPs originate at larger radii where the electron density is lower. In projection on the sky, this results in a series of concentric, disk-like emission regions, with smaller angular extents corresponding to larger ALP masses.

For spatially resolved clusters, this mass–radius mapping provides a powerful handle to isolate the ALP-induced signals and to determine the coupling strength as a function of mass. In realistic systems, departures from spherical symmetry and substructure will distort these idealized circular patterns, but the underlying radial dependence remains a robust qualitative feature of the signal morphology.

\section{Partial-sky Power spectrum and Mode coupling}
\label{sec:partial_sky}

Galaxy clusters subtend relatively small angular scales on the sky, and consequently the signals associated with them are confined to localized sky regions. As a result, the analysis must be performed on a partial sky rather than over the full celestial sphere. In such cases, the statistical properties of the estimated power spectrum differ from those obtained under full-sky coverage, as the orthogonality of spherical harmonics is broken by the finite sky mask.

Let the observed sky be represented by a window function $W(\hat{n})$ that takes the value unity inside the observed region and zero elsewhere. The observed sky map is then given by the product of the true sky signal and this window function. In harmonic space, this multiplication leads to a convolution, such that the measured (pseudo-)power spectrum $\bar{C}_\ell$ is related to the true underlying power spectrum $C_\ell$ via
\begin{equation}
\bar{C}_{\ell} = \sum_{\ell'} M_{\ell\ell'} \left( B_{\ell'}^{2} C_{\ell'} + N_{\ell'} \right),
\label{eq:modecouple}
\end{equation}
where $B_{\ell}$ denotes the instrumental beam transfer function and $N_{\ell}$ represents the noise power spectrum. The matrix $M_{\ell\ell'}$ encodes the mode coupling induced by the partial sky coverage.

The mode-coupling matrix depends solely on the geometry of the observed region and can be expressed as \cite{Hivon_2002}
\begin{equation}
M_{\ell_1 \ell_2}
=
\frac{2\ell_2 + 1}{4\pi}
\sum_{\ell_3}
W_{\ell_3}
\begin{pmatrix}
\ell_1 & \ell_2 & \ell_3 \\
0 & 0 & 0
\end{pmatrix}^{2},
\label{eq:maskkernel}
\end{equation}
where $W_{\ell}$ denotes the spherical harmonic coefficients of the window function and the Wigner $3j$ symbol enforces the angular momentum coupling conditions arising from the partial sky coverage.

The presence of the window function leads to mixing between different multipoles, such that power at a given $\ell$ receives contributions from a range of true multipoles $\ell'$. For a binary mask corresponding to a localized sky region, this coupling becomes particularly significant at high multipoles, where the effective angular resolution is most sensitive to the mask geometry.

In the present analysis, the noise is assumed to be Gaussian and spatially uncorrelated, allowing the noise power spectrum to be treated as diagonal in harmonic space. The effect of partial sky coverage on the noise contribution is therefore captured through the same mode-coupling formalism. Rather than attempting to invert the coupling matrix to recover a full-sky estimator, we explicitly retain the convolution structure in Eq.~\eqref{eq:modecouple} and perform all likelihood evaluations using the pseudo-$C_\ell$ formalism. This approach enables a consistent treatment of partial-sky observations and naturally incorporates the impact of sky masking in the statistical inference of the underlying signal.

\acknowledgments
    This work is a part of the $\langle \texttt{data|theory}\rangle$ \texttt{Universe-Lab}, supported by the TIFR  and the Department of Atomic Energy, Government of India. The authors express their gratitude to the $\langle \texttt{data|theory}\rangle$ \texttt{Universe-Lab} and the TIFR CCHPC facility for meeting the computational needs. 
 Also, the following packages were used for this work: Astropy \cite{astropy:2013,astropy:2022,astropy:2018},
, NumPy \cite{harris2020array}
CAMB \cite{2011ascl.soft02026L}, SciPy \cite{2020SciPy-NMeth}, SymPy \cite{10.7717/peerj-cs.103}, Matplotlib \cite{Hunter:2007}, emcee \cite{Foreman_Mackey_2013}, HEALPix (Hierarchical Equal Area isoLatitude Pixelation of a sphere)\footnote{Link to the HEALPix website http://healpix.sourceforge.io}\cite{2005ApJ...622..759G,Zonca2019} and PySM \cite{Thorne_2017}.

\bibliography{references.bib}








\end{document}